%% file: StaticGravity.tex
\documentclass[10pt,a4paper,twocolumn]{article}
\usepackage[hmargin=17mm,vmargin=19mm,columnsep=6mm]{geometry}

\usepackage[T1]{fontenc}
\usepackage{amsmath}
\usepackage{newtxtext,newtxmath}
\usepackage[scaled=0.92]{helvet}
\usepackage{microtype,graphicx}
\usepackage{titlesec,titling,xcolor}

\definecolor{uclblue}{RGB}{25,54,98}
\definecolor{desycyan}{rgb}{0.00,0.68,0.93}
\definecolor{desyorange}{rgb}{0.96,0.52,0.07}

\usepackage{cite}
\usepackage[colorlinks=true,linkcolor=desycyan,citecolor=desycyan,urlcolor=desycyan]{hyperref}
\usepackage[font=small,labelfont=bf,labelsep=period]{caption}
\newcommand{\Gb}{\bar\Gamma}
\newcommand{\Wc}{\mathcal W}
\newcommand{\dd}{\mathrm d}
\newcommand{\calH}{\mathcal H}
\DeclareMathOperator{\artanh}{artanh}
\DeclareMathOperator{\DR}{DR}
\DeclareMathOperator{\Pf}{Pf}
\DeclareMathOperator{\FP}{FP}
\DeclareMathOperator{\Res}{Res}
\DeclareMathOperator{\ext}{ext}

\titleformat{\section}{\sffamily\bfseries\large}{}{0pt}{}
\titleformat{\subsection}{\sffamily\bfseries\fontsize{10.5}{12.5}\selectfont}{}{0pt}{}
\titleformat{\paragraph}[runin]{\sffamily\bfseries\small}{}{0pt}{}[.]
\titlespacing*{\section}{0pt}{12pt plus 2pt}{5pt}
\titlespacing*{\subsection}{0pt}{9pt plus 2pt}{3pt}
\titlespacing*{\paragraph}{0pt}{4pt plus 1pt}{0.6em}
\pretitle{\noindent\sffamily\bfseries\fontsize{24}{27}\selectfont}
\posttitle{\par\vspace{12pt}}
\preauthor{\noindent}
\postauthor{\par\vspace{4pt}}
\predate{}\postdate{}
\begin{document}
\title{\LARGE Gravity between two masses beyond Newton's law}

\author{{Zhengwen Liu}${}^1$ and {Xing-Bo Yuan}${}^2$}
\date{}

\twocolumn[{
\maketitle
\vspace{-1.5em}
\noindent${}^1$\emph{\small School of Physics \& Shing-Tung Yau Center, Southeast University, Nanjing 210096, China}\par
\noindent${}^2$\emph{\small Institute of Particle Physics \& Key Laboratory of Quark and Lepton Physics (MOE), Central China Normal University, Wuhan 430079, China}\par
\vspace{15pt}

\noindent{%
Newton's law of universal gravitation describes the attraction between two masses through an inverse-distance potential~\cite{Newton:1687principia}. In Einstein's theory gravity gravitates: each mass reshapes the spacetime geometry through which the other acts \cite{Einstein:1915field,Einstein:1916foundation}. More than a century later, extracting the two-body interaction from this coupled geometry remains a fundamental analytical challenge. Even without motion or radiation, the static sector of the interaction harbours an infinite hierarchy of field couplings: {\it the purest laboratory for gravitational non-linearity}. Here we determine the static interaction from ninth to twelfth order in Newton's constant for arbitrary mass ratios, and derive a closed expression for the first backreaction beyond the Schwarzschild test-body limit to all orders in the gravitational coupling. We achieve both advances by formulating the problem around the stationary interaction energy. An exact one-body background resums one source's self-interaction, stationarity reduces the fields required from the other, and integrated response identities reveal cancellations across the coupled modes. Complementary expansions then resolve the complete mass dependence at finite order. Together, these results bridge the gap between an infinitesimal probe and comparable masses, exposing hidden analytical simplicity within Einstein's theory and providing essential building blocks for modelling orbital dynamics and gravitational-wave generation.}
\vskip 20pt
}]

\begin{figure*}[!t]
 \centering\includegraphics[width=\textwidth]{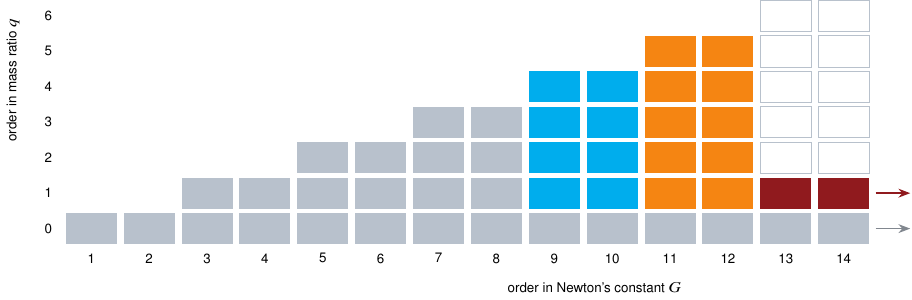}
 \caption{\textbf{Known and new terms of the static interaction.}
 Each cell is one term of the potential between two masses. Moving right raises the power of Newton's constant $G$, and moving up the power of the mass ratio: the two axes are the two expansions of the interaction. Terms that differ only by exchanging the bodies appear once. Grey: known before this work. Colour: obtained here, namely all terms beyond the probe term at $G^9$ and $G^{10}$ (cyan) and at $G^{11}$ and $G^{12}$ (orange), and the first backreaction, the second row, at every higher order (maroon). Arrows mark the two rows now known to all orders; empty cells remain open.}
 \label{fig:coverage}
\end{figure*}

\noindent
In 1687, Newton formulated the gravitational interaction between two bodies into a universal law~\cite{Newton:1687principia}, expressed in modern notation through the potential
\begin{equation}
V_{\rm N}(r)=-\frac{Gm_1m_2}{r}.
 \label{eq:newton}
\end{equation}
This single inverse-distance law linked terrestrial motion to the architecture of the heavens. Its simplicity became the starting
point for understanding gravitation.

\vskip 2pt
Einstein fundamentally recast the nature of gravity~\cite{Einstein:1915field,Einstein:1916foundation}. Mass curves spacetime, and the geometry that mediates the interaction responds to the gravitational field itself. Linear superposition no longer holds: each mass warps the spacetime through which the other interacts, so that gravity acts as a source for itself. Newton's single term becomes the leading piece of an infinite nonlinear structure.

\vskip 2pt
This nonlinearity is present even before either body moves or radiates. The static sector, the part of the interaction independent of velocities, isolates gravitational nonlinearity in its purest form, distilling a foundational question at the heart of Einstein's theory:~how two masses interact through the nonlinear structure of spacetime geometry alone.

\vskip 2pt
More than a century of analytical work has progressively unraveled this structure. Early calculations by de Sitter and by Lorentz and Droste extended relativistic mechanics to interacting bodies~\cite{deSitter:1916second,LorentzDroste:1917II}. In 1938, Einstein, Infeld and Hoffmann derived the leading relativistic equations of motion directly from the vacuum field equations~\cite{Einstein:1938yz}, including a static interaction term quadratic in Newton's constant $G$. Post-Newtonian theory then organized the slow-motion, weak-field problem into a systematic expansion, through which Damour, Sch\"afer, Blanchet and their collaborators established the dynamics to high orders alongside the regularization methods required to define point-mass sources~\cite{DamourDeruelle:1985,DamourSchafer:1985,DamourJaranowskiSchafer:2001,BlanchetDamourEspositoFarese:2004,DamourJaranowskiSchafer:2014}.

\vskip 2pt
Gravitational-wave astronomy has given this long pursuit direct observational urgency. Waves from merging black holes and neutron stars carry the imprint of two-body dynamics across the Universe to our interferometers~\cite{AbbottEtAl:2016GW150914,AbbottEtAl:2017GW170817}. During inspiral, their evolving phase encodes both the conservative interaction governing the orbit and the energy carried away by radiation. Extracting physics encoded in that phase requires precise theoretical predictions for both, providing a central motivation for high-order calculations~\cite{Cutler:1993zz,Buonanno:1998gg,Blanchet:2023sbv}.

\vskip 2pt
Quantum field theory transformed the tools available for this endeavour. Early diagrammatic treatments recovered known low-order potentials and the Schwarzschild geometry~\cite{Iwasaki:1971,Duff:1973}. Modern effective field theory organized field exchanges and nonlinear couplings through systematic power counting~\cite{Goldberger:2004jt,Porto:2016,Dlapa:2023hsl}, while S-matrix methods extracted classical physical quantities from two-body scattering amplitudes~\cite{Bjerrum-Bohr:2018xdl,CheungRothsteinSolon:2018}. These approaches brought advanced Feynman-integral techniques—originally forged for particle physics—to gravitational physics.
In the static sector, four- and five-loop calculations reached $G^5$ and $G^6$~\cite{Foffa:2016rgu,DamourJaranowski:2017,Foffa:2019hrb,Blumlein:2019zku}; six-loop integration subsequently reached $G^7$, and the reconstruction of correlation functions completed $G^8$, with full mass dependence~\cite{Brunello:2025gpf,Brunello:2026anu}. These milestones demonstrate the reach of modern multiloop methods, but also the formidable cost of advancing one power of $G$ at a time.

\vskip 2pt
There is another route into the problem. If a probe of mass $m_2$ is too small to alter the geometry produced by a mass $m_1$, the Schwarzschild solution gives its static interaction in closed form~\cite{Schwarzschild:1916uq,Foffa:2019hrb}. In the harmonic radial coordinate $r$, with the dimensionless coupling $\alpha=Gm_1/(rc^2)$ and the probe's rest energy subtracted, the result is
\begin{equation}
 \frac{V_{\rm S}}{m_2c^2}=\Phi_0(\alpha)
 =\sqrt{\frac{1-\alpha}{1+\alpha}}-1.
 \label{eq:testbody}
\end{equation}
An infinite series in $G$ is thus resummed into a single function. Its leading term, $-\alpha$, recovers Newton's potential \eqref{eq:newton}; every higher power is fixed by the same square root. The probe samples this nonlinear geometry without changing it. The natural question is how this exact nonlinear geometry responds when the smaller mass contributes its own field. Expanding in the mass ratio captures it order by order, the organizing principle of gravitational self-force theory~\cite{MinoSasakiTanaka:1997,QuinnWald:1997}.

\vskip 4pt
Here we advance the finite-order and exact-background descriptions together. We determine the complete static interaction between two point masses through $G^{12}$, four orders beyond the previous frontier and with the full dependence on the mass ratio, and derive its leading backreaction as a closed function of the coupling. 
Both follow from the stationary energy of the gravitational field, where variational stationarity isolates the minimal field information required for the interaction energy. Complementary field and spectral calculations expose a collective cancellation at first order in the mass ratio and reduce the higher-order responses required to complete the four new orders.
The two results overlap, as illustrated in Figure~\ref{fig:coverage}, and complement each other: one is complete in the coupling at first order in the mass ratio, the other complete in the mass ratio at each order in the coupling. 

\begin{figure*}[!t]
 \centering\includegraphics[width=\textwidth,trim=0 6 0 0,clip]{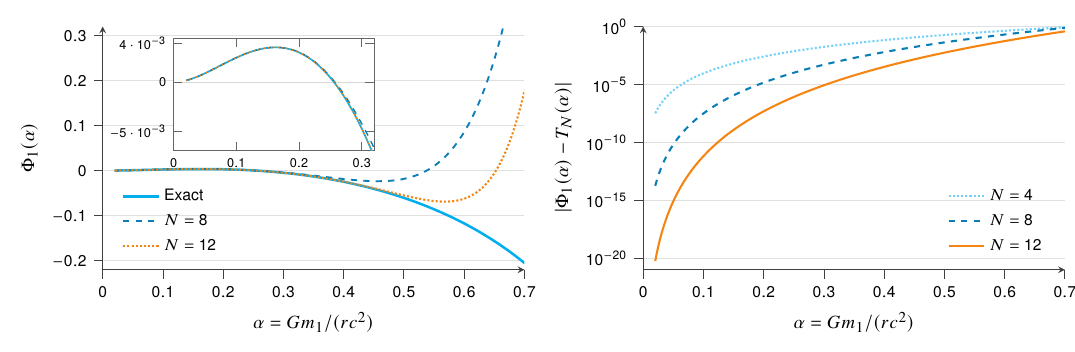}
 \caption{\textbf{The first gravitational backreaction to all orders in $G$.}
 \textit{Left}. The exact closed-form function $\Phi_1$ (cyan) and its weak-field expansions through $G^8$ (blue, dashed) and $G^{12}$ (orange, dotted); the inset magnifies the zero crossing at $\alpha\simeq0.256$.
 \textit{Right}. Absolute truncation errors $|\Phi_1-T_N|$ of the partial sums $T_N$ through $G^N$, for $N=4$ (light cyan, dotted), $8$ (blue, dashed) and $12$ (orange, solid). All curves follow exactly from the closed function, without fitting, within its convergence $|\alpha|<1$.}
 \label{fig:resummation}
\end{figure*}

\subsection{The first backreaction in closed form}
The probe's own gravity first alters the interaction at second order in its mass. This is the first-order self-force (1SF) contribution, whereas the Schwarzschild probe term is the zeroth (0SF). Labelling the bodies so that $m_2\le m_1$, we expand $V/(m_2c^2)=\Phi_0(\alpha)+q\Phi_1(\alpha)+O(q^2)$ with $q=m_2/m_1$. We find
\begin{equation}
 \boxed{\begin{aligned}
 \Phi_1(\alpha)&=\tfrac12\bigl(\alpha\Phi_0'\bigr)^2\bigl[1-4\alpha+\tfrac43\alpha^2\artanh\alpha\bigr]\\
 &=\frac{\alpha^2\bigl(3-12\alpha+4\alpha^2\artanh\alpha\bigr)}{6(1-\alpha)(1+\alpha)^3},
 \end{aligned}}
 \label{eq:physical}
\end{equation}
where $\alpha\Phi_0'=-\alpha(1+\alpha)^{-3/2}(1-\alpha)^{-1/2}$ is the scaled probe force and $\artanh\alpha=\frac12\log[(1+\alpha)/(1-\alpha)]$. Both $\Phi_0$ and $\Phi_1$ are real for $0\le\alpha<1$ and analytic for $|\alpha|<1$.

\vskip 2pt
A static mass sources gravity with an effective strength equal to its physical mass times its redshift factor, the rate at which its proper time runs relative to coordinate time. Holding the physical mass $m_1$ fixed therefore does not fix its strength as a source: the gravitational field of $m_2$ slows proper time at $m_1$, which directly accounts for the unit term inside the brackets in the first line of equation~\eqref{eq:physical}. 
The remaining terms arise from the quadratic field response—the energy of the gravitational field's response to $m_2$ in the exact Schwarzschild geometry of $m_1$. Both contributions are quadratic in the effective strength of the second body, so the prefactor contains the square of its redshift factor, $(1+\Phi_0)^2=(1-\alpha)/(1+\alpha)$. The weak-field expansion $\Phi_1=\frac12\alpha^2-3\alpha^3+6\alpha^4-\frac{31}{3}\alpha^5+\cdots$ reproduces every known coefficient of $G^nm_1^{n-1}m_2^2$ through $G^8$ and supplies all others. 
Because these Taylor coefficients are built from harmonic numbers, equation~\eqref{eq:physical} also shows that all static 1SF coefficients are purely rational, free of transcendental constants, at every order in $G$.

\vskip 4pt
We prove equation~\eqref{eq:physical} to all orders in $G$ within dimensionally regulated perturbation theory, with space continued to $d=3+\epsilon$ dimensions. The derivation decomposes the quadratic response into angular modes (multipoles) about the heavier mass and hinges on a collective cancellation. At each order in the coupling, the response in multipole $\ell$ is a rational function of $\ell$. Subtracting its large-$\ell$ limit leaves the regulated sum unchanged, and the remainder, continued to every $\ell$, sums to zero by an exact identity of the coupled radial equations. Only the monopole and the dipole, which carry fewer tensor harmonics, depart from this generic continuation, and these departures alone give the quadratic response in closed form. The perturbations keep a complicated angular structure, but their integrated effect is simple.
Multiple independent calculations also confirm this closed-form expression at every order tested. Four independently coded mode-sum computations reproduce its coefficients, exactly through $G^{50}$ in the longest run. Moreover, a momentum-space Born series of flat-space Feynman integrals, without mode sums, gives those at $G^9$ and $G^{11}$, and exact spectral sums give those at $G^7$, $G^9$ and $G^{11}$.

\vskip 4pt
We work in harmonic gauge, implemented through a gauge-fixing term in the action, with the two masses held at fixed positions.  Recent effective-field-theory calculations~\cite{Brunello:2025gpf,Brunello:2026anu} adopt the same gauge, allowing direct term-by-term comparison. A potential written in terms of the coordinate distance $r$ depends on the gauge, even in the probe limit. In the standard Schwarzschild coordinates, with radial coordinate $R=r+Gm_1/c^2$, the probe energy~\eqref{eq:testbody} takes the familiar form $\sqrt{1-2Gm_1/(Rc^2)}-1$.
Although the static potential itself is gauge-dependent, combined with the velocity-dependent and hereditary contributions in the same gauge, it determines gauge-invariant quantities, such as the binding energy of a circular orbit as a function of orbital frequency.

\vskip 4pt
The closed form \eqref{eq:physical} also shows how the first backreaction changes from weak to strong fields. In weak fields the unit term dominates, and $\Phi_1$ is a small positive correction captured by its first few terms. The zero crossing at $\alpha\simeq0.256$ (Figure~\ref{fig:resummation}, {\it left}) reflects a cancellation within the 1SF contribution, between the unit term and the quadratic response; the probe term and the higher backreactions still contribute there. Because the weak-field coefficients alternate in sign and grow, truncations lose accuracy as $\alpha$ increases (Figure~\ref{fig:resummation}, {\it right}); at $\alpha=0.5$, even the expansion through $G^{12}$ misses $\Phi_1$ by 9\%. The exact function $\Phi_1$ eliminates this truncation error at first order in the mass ratio.

\begin{figure*}[!t]
{\setlength{\medmuskip}{3mu plus 2mu minus 1mu}%
\begin{align}
V={}&{-\frac{Gm_1m_2}{2r}}\nonumber\\[2pt]
&{+\frac{G^2m_1^2m_2}{2r^2c^2}}\nonumber\\[2pt]
&{-\frac{G^{3}}{r^{3}c^{4}}\Bigl[\frac{1}{2}m_1^{3}m_2 + \frac{3}{2}m_1^{2}m_2^{2}\Bigr]}\nonumber\\[2pt]
&{+\frac{G^{4}}{r^{4}c^{6}}\Bigl[\frac{3}{8}m_1^{4}m_2 + 6m_1^{3}m_2^{2}\Bigr]}\nonumber\\[2pt]
&{-\frac{G^{5}}{r^{5}c^{8}}\Bigl[\frac{3}{8}m_1^{5}m_2 + \frac{31}{3}m_1^{4}m_2^{2} + \frac{141}{8}m_1^{3}m_2^{3}\Bigr]}\nonumber\\[2pt]
&{+\frac{G^{6}}{r^{6}c^{10}}\Bigl[\frac{5}{16}m_1^{6}m_2 + \frac{91}{6}m_1^{5}m_2^{2} + \frac{653}{6}m_1^{4}m_2^{3}\Bigr]}\nonumber\\[2pt]
&{-\frac{G^{7}}{r^{7}c^{12}}\Bigl[\frac{5}{16}m_1^{7}m_2 + \frac{190}{9}m_1^{6}m_2^{2} + \frac{37651}{144}m_1^{5}m_2^{3} + \frac{2926}{9}m_1^{4}m_2^{4}\Bigr]}\nonumber\\[2pt]
&{+\frac{G^{8}}{r^{8}c^{14}}\Bigl[\frac{35}{128}m_1^{8}m_2 + \frac{248}{9}m_1^{7}m_2^{2} + \frac{1059}{2}m_1^{6}m_2^{3} + \frac{89383}{36}m_1^{5}m_2^{4}\Bigr]}\nonumber\\[2pt]
&-\frac{G^{9}}{r^{9}c^{16}}\Bigl[\frac{35}{128}m_1^{9}m_2{\color{desycyan}{}+ \frac{1574}{45}m_1^{8}m_2^{2} + \frac{1387541}{1440}m_1^{7}m_2^{3} + \frac{110528}{15}m_1^{6}m_2^{4} + \frac{8713729}{1152}m_1^{5}m_2^{5}}\Bigr]\nonumber\\[2pt]
&+\frac{G^{10}}{r^{10}c^{18}}\Bigl[\frac{63}{256}m_1^{10}m_2{\color{desycyan}{}+ \frac{429}{10}m_1^{9}m_2^{2} + \frac{129269}{80}m_1^{8}m_2^{3} + \frac{1655129}{90}m_1^{7}m_2^{4} + \frac{15680393}{240}m_1^{6}m_2^{5}}\Bigr]\nonumber\\[2pt]
&-\frac{G^{11}}{r^{11}c^{20}}\Bigl[\frac{63}{256}m_1^{11}m_2{\color{desyorange}{}+ \frac{5429}{105}m_1^{10}m_2^{2} + \frac{9263679437}{3628800}m_1^{9}m_2^{3} + \frac{16446596}{405}m_1^{8}m_2^{4} + \frac{3868176257}{17280}m_1^{7}m_2^{5} + \frac{5468741}{27}m_1^{6}m_2^{6}}\Bigr]\nonumber\\[2pt]
&+\frac{G^{12}}{r^{12}c^{22}}\Bigl[\frac{231}{1024}m_1^{12}m_2{\color{desyorange}{}+ \frac{6406}{105}m_1^{11}m_2^{2} + \frac{129106847}{33600}m_1^{10}m_2^{3} + \frac{9262613639}{113400}m_1^{9}m_2^{4} + \frac{8404720391}{12960}m_1^{8}m_2^{5} + \frac{511671503}{270}m_1^{7}m_2^{6}}\Bigr]\nonumber\\[2pt]
&+(m_1\leftrightarrow m_2) \,+\, O(G^{13})\label{eq:fullpotential}
\end{align}
}
 \vspace{-15pt}
 \caption*{\textbf{The static gravitational potential through twelfth order in $G$.}
The exchange $m_1\leftrightarrow m_2$ applies to the entire displayed sum, so terms symmetric under exchange appear with half their coefficient. The first two lines recover Newton's potential and the static Einstein-Infeld-Hoffmann term in harmonic coordinates. Black: terms known prior to this work, comprising the complete static interaction through $G^8$ and the test-body (0SF) terms to all orders. Cyan and orange: the new contributions at $G^9$--$G^{10}$ and $G^{11}$--$G^{12}$. At fixed $m_1$, the terms linear and quadratic in $m_2$ correspond to the Taylor expansions of the exact 0SF and 1SF functions.}
\end{figure*}

\subsection{Twelve orders in Newton's constant}
The probe term and the leading backreaction, together with exchange symmetry, determine the complete static potential through $G^4$.
From $G^5$ onward, terms appear in which both masses are at least cubed, giving the second-order (2SF) and higher self-force contributions.
In the expansion about one body's background, these involve the cubic and higher responses of the field to the other body.
For comparable masses, their importance cannot be assessed by a small-mass-ratio expansion alone.

\vskip 4pt
Reaching the four new orders requires independent self-force information through 4SF at $G^9$--$G^{10}$ and through 5SF at $G^{11}$--$G^{12}$ (Figure\,\ref{fig:coverage}).
We determine these contributions by expanding in $G$ about flat space while retaining the complete dependence on both masses. 
Through $G^{12}$, the entire static potential is fixed by 21 independent connected field correlators, described below (see Supplementary Material).
Eleven of them are fixed by the closed probe and first-backreaction functions; three further correlators were known from earlier work~\cite{Foffa:2016rgu,DamourJaranowski:2017,Brunello:2025gpf,Brunello:2026anu}.
All but three of the correlators are obtained analytically, and these determine $G^9$ and $G^{10}$ completely.
The remaining three first enter at $G^{11}$ in terms where both masses appear at least quartically (3SF and higher).
We compute these three correlators numerically to roughly 22 significant digits and identify them with rational numbers, which they match to relative differences below $4\times10^{-23}$ (empirical uncertainties below $8\times10^{-23}$), their analytic evaluation is the subject of ongoing work.

\vskip 4pt
Equation~\eqref{eq:fullpotential} presents our result.
We observe that at every order $G^n$, all mass monomials carry the same sign $(-1)^n$, and within each order the coefficients increase towards terms with more nearly balanced powers of the two masses.
At $G^{12}$, the coefficient rises from $231/1024\simeq0.23$ for $m_1^{12}m_2$ to approximately $1.9\times10^6$ for $m_1^7m_2^6$, a span of nearly seven orders of magnitude.
For equal masses every monomial at a given order takes the same value, so the twelfth-order term is dominated by the higher backreactions.

\subsection{From probe to equal masses}
The complete mass dependence allows us to assess how much of the static interaction is captured by the probe term and its first backreaction as the masses become comparable.
To quantify this distinction, we compare mass ratios at fixed total mass $M=m_1+m_2$ and coordinate separation $r$. We use the reduced mass $\mu=m_1m_2/M$, the symmetric mass ratio $\nu=\mu/M=q/(1+q)^2$ and the coupling $u=GM/(rc^2)=(1+q)\alpha$. The potential becomes $V/(\mu c^2)=\sum_n u^n\mathscr P_n(\nu)$, with $\nu$ ranging from zero in the probe limit to $1/4$ for equal masses. The $\nu^0$ and $\nu^1$ contributions—corresponding to the 0SF and 1SF sectors in these variables are the closed-form functions $\Phi_0(u)$ and $\Phi_1^{\rm sym}(u)=\Phi_0(u)-u\Phi_0'(u)+\Phi_1(u)$. Here, the first two terms of $\Phi_1^{\rm sym}$ merely re-express the probe energy in the new variables, while $\Phi_1$ adds the leading backreaction. Higher powers of $\nu$ encode the higher-order backreactions, now known through $G^{12}$.

\begin{figure*}[t]
 \centering\includegraphics[width=\textwidth,trim=0 6 0 0,clip]{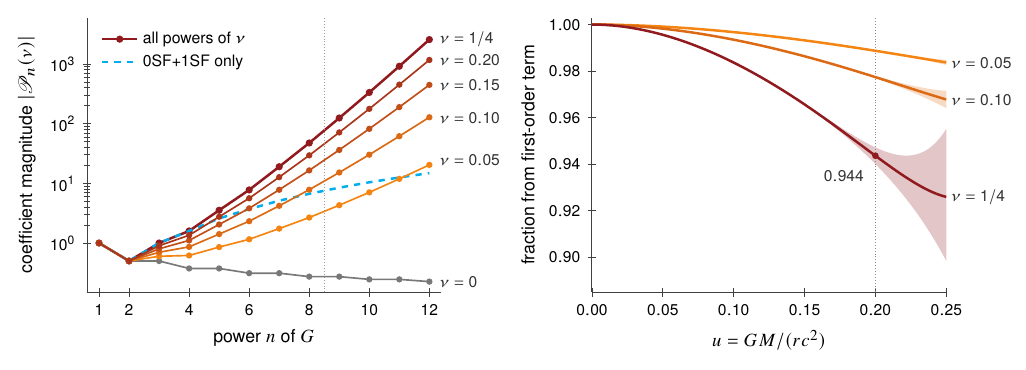}
 \caption{\textbf{How backreactions build the static interaction.}
 \textit{Left}. Magnitudes $|\mathscr P_n(\nu)|$ of the coefficients of $u^n$, from the probe limit ($\nu=0$, grey) to equal masses ($\nu=\frac14$), with mass ratios labelled at the curve ends. Dashed: the $\nu^0$ and $\nu^1$ (0SF and 1SF) terms alone, at equal masses. The dotted line separates the new orders.
 \textit{Right}. Fraction of the correction beyond the probe term, $V-\mu c^2\Phi_0(u)$, carried by the first-order term $\nu\mu c^2\Phi_1^{\rm sym}(u)$, using the closed functions and the higher powers of $\nu$ through $G^{12}$. Shading varies the denominator by the last included higher-backreaction term, as a truncation diagnostic rather than a rigorous bound. Dotted: $u=0.2$, a reference value, corresponding to $r=5GM/c^2$, the harmonic radius of the innermost stable circular orbit of a test body around a mass $M$.}
 \label{fig:mutual}
\end{figure*}

\vskip 4pt
Higher-order backreactions dominate the high-order coefficients (Figure\,\ref{fig:mutual}, {\it left}). In the probe limit, coefficients remain of order unity, as required by the square-root branch point of equation~\eqref{eq:testbody} at $u=\pm1$. For comparable masses, however, they exhibit approximately geometric growth with alternating signs; at equal masses, the ratio of successive coefficients rises from $2.2$ at $G^6$ to $2.8$ at $G^{12}$. At $G^{12}$, the equal-mass coefficient is about $2.57\times10^3$, to which the 0SF and 1SF terms contribute only $0.58\%$, whereas higher backreactions account for the remainder, with the $\nu^3$ and $\nu^4$ powers alone providing $79\%$. 
Within the $G^{12}$ coefficient, terms quadratic and higher in $\nu$ exceed the combined 0SF and 1SF contribution for $\nu\gtrsim0.017$.

\vskip 4pt
A large coefficient does not necessarily imply a dominant contribution, because it multiplies a high power of the coupling. At $u=0.2$, the alternating terms decrease: for equal masses, the $G^{12}$ contribution is about $0.56$ times that at $G^{11}$. The first-order (1SF) term captures the most majority of the correction beyond the probe potential (Figure\,\ref{fig:mutual}, {\it right}), with higher backreactions contributing only a few percent for equal masses. 

\vskip 4pt
Together, the closed-form function and the finite-order expansion disentangle two distinct questions that neither answers alone: how much is lost by truncating the weak-field expansion, and how much by truncating the mass-ratio expansion.

\subsection{The potential from the stationary energy}
Our calculations separate the self-interaction of the gravitational field from the nonlinear coupling of each point mass to it. We replace each particle's coupling with an auxiliary linear source, compute the stationary energy of the gravitational field, and restore the physical couplings algebraically at the end. The resulting energy admits two complementary expansions. Keeping the exact field of one body as the background sums all powers of its source~\cite{CheungEtAl:2024EMR,KosmopoulosSolon:2024}. 
With the couplings restored, the terms linear and quadratic in the other mass give the probe energy and the leading backreaction, \eqref{eq:testbody} and \eqref{eq:physical}, and higher powers generate successive orders in the mass ratio to all orders in $G$. Expanding in both sources about flat space instead resolves every distribution of the masses at a fixed order in $G$. Where they overlap, the two expansions compute the same coefficients in different ways and so check each other.

\vskip 4pt
The energy with linear sources is generated by dimensionless connected correlators $\Gb_{a,b}=\Gb_{b,a}$, which join $a$ insertions of the first source to $b$ of the second and first contribute at $G^{a+b-1}$; an exact, finite transformation then fixes the source strengths in terms of the physical masses, restoring the particle couplings without new spatial integrations~\cite{Brunello:2026anu}. A reflection symmetry of the static field equations~\cite{Buchdahl:1959} makes the correlators vanish unless $a+b$ is even. Consequently, the new integrals at $G^9$ and $G^{11}$ also determine $G^{10}$ and $G^{12}$~\cite{Foffa:2019hrb,Brunello:2026anu}. Through $G^{12}$ the potential rests on 21 correlators: the probe and first-backreaction towers, three known from earlier work~\cite{Foffa:2016rgu,DamourJaranowski:2017,Brunello:2025gpf,Brunello:2026anu} and seven new ones.

\vskip 4pt
Although it is well established that tree-level diagrams sum to the classical field energy~\cite{BoulwareBrown:1968}, the decisive simplification comes from the variational stationarity of this energy.  Because a variation of the fields about their classical solution alters the energy only at second order, field responses known through order $p$ in a source determine the interaction energy through order $2p+1$~\cite{Gonze:1995variational}---a principle mirrored in the Fokker-action approach to post-Newtonian dynamics~\cite{BernardEtAl:2016}. Crucially, we enforce stationarity before integrating, eliminating the highest-order field responses at the outset rather than solving for them only to have them cancel at the end.

\vskip 4pt
A diagrammatic count illustrates the scale of this reduction. 
The connected diagrams joining the two sources are trees of scalar and metric propagators that correspond to multiloop momentum-space integrals, reaching eight loops at $G^9$ and ten loops at $G^{11}$. 
Up to exchange of the bodies, there are 282 such diagrams at $G^7$, 4,235 at $G^9$, and 74,092 at $G^{11}$. Sub-trees shared across diagrams collect into reusable field responses expanded in powers of the sources: the Newtonian scalar potential, the spatial metric it sources, and so on. Combining these responses into {\it pairings} reproduces every diagram with its symmetry factor, with the highest-order field entering through a single quadratic term. For instance, the 74,092 diagrams at $G^{11}$ collapse into just 18 pairings of six field responses. Although each pairing still involves nontrivial spatial integrals, vast amounts of redundant field information are bypassed: only the integrated pairing of each response is required, and many individual field components never need to be solved for.
 
\vskip 4pt
Point masses make the fields singular, and the energy is regularized dimensionally in $d=3+\epsilon$ spatial dimensions. The poles in $\epsilon$ cancel in the total energy but leave finite local terms at the particles, sensitive to the $\epsilon$-dependence of the fields. Bulk integrals, boundary terms, and these local contributions form a single regularized quantity: omitting one changes the potential. At $G^{11}$ the result also depends on how the regular part of a field produced by both bodies varies with the dimension, information that no three-dimensional solution contains. We isolate this dimensional dependence field by field and resolve it using exact $d$-dimensional data.

\vskip 4pt
Every correlator is evaluated in at least two independent ways. Exact evaluations, mainly an analytic reduction of the pairings, cover all but $\Gb_{4,8}$, $\Gb_{5,7}$ and $\Gb_{6,6}$; numerical ones cover all. Sphere by sphere around one body, they reproduce the exact correlators of $G^9$ and $\Gb_{3,9}$ to at least 19 significant digits and give the other three to about 22; on a grid adapted to both bodies, with the same exact $d$-dimensional data for one two-body field, they confirm these to about 12 digits.

\vskip 4pt
Rational values emerge from the analytic evaluations without a rational ansatz.
At $G^9$, intermediate expressions contain multiple polylogarithms of weight up to five, iterated integrals of modular forms and constants such as $\pi^2$, $\zeta(3)$ and Catalan's constant, all of which cancel in the complete correlators. Some cancel only once bulk, boundary and local pieces are combined: in $\Gb_{5,5}$, for example, a $\pi^2$ term of the three-dimensional integral is cancelled exactly by the local correction.

\vskip 4pt
The new values also distinguish rationality from the stronger integer pattern of the lower orders. Every nonvanishing correlator through $G^9$ is a positive integer in the stated normalization. At $G^{11}$, the analytic $\Gb_{3,9}=1770222336/5$ breaks this pattern, while $\Gb_{4,8}\simeq5675220992/3$ supports a second noninteger value numerically. Every coefficient determined analytically remains rational, including the entire first-backreaction series. For the three numerical inputs, the quoted rational values remain identifications awaiting analytic evaluation. Whether rationality persists at higher orders remains open.

\vskip 4pt
A more detailed account of the derivations and evaluations is given in the Supplementary Material, to be given in full in \cite{ZCXB:inPreparation}.

\subsection{Outlook}
In summary, more than three centuries after Newton's universal law, the static potential between two masses in Einstein's theory is now known to all orders in $G$ at first order in the mass ratio, and through twelfth order in $G$ for arbitrary masses.

\vskip 4pt
The framework developed here is not tied to the static limit: variational stationarity, field-response pairings, and exact one-body backgrounds, can also be used for moving sources, which could open a route to the velocity-dependent sectors.
Combined with those sectors, our static results supply direct inputs to post-Newtonian bound dynamics~\cite{FoffaPortoRothsteinSturani:2019,Blanchet:2024LRR,PortoRiva:2026}, effective-one-body models~\cite{Buonanno:1998gg}, and, with the radiated flux~\cite{Blanchet:2023sbv}, gravitational-wave phasing~\cite{BlanchetEtAl:2023phasing}. The response formulation also suggests a route to post-Minkowskian scattering~\cite{Damour:2016PM,BernEtAl:2019PM,KalinLiuPorto:2020,DlapaEtAl:2022PM,BernEtAl:2022PM,Dlapa:2024cje,DriesseEtAl:2025PM,DlapaEtAl:2025PM5,Dlapa:2026oyq,BernEtAl:2025G5}, provided time-dependent propagation, recoil and tails are included; whether the collective cancellation survives once the sources move remains open.

\clearpage
\onecolumn
\section*{Supplementary Material}
\pdfbookmark[1]{Supplementary Material}{supplementarymaterial}
\label{meth:section}
\setcounter{equation}{0}
\renewcommand{\theequation}{M\arabic{equation}}
\renewcommand{\theHequation}{M.\arabic{equation}}
\setcounter{figure}{0}
\renewcommand{\thefigure}{M\arabic{figure}}
\renewcommand{\theHfigure}{M.\arabic{figure}}

This Supplementary Material explains how the two main results in this work are obtained: the first backreaction in closed form, in equation~\eqref{eq:physical}, and the complete potential through $G^{12}$, in equation~\eqref{eq:fullpotential}. Both come from one construction, which separates the nonlinear self-interaction of the gravitational field from the nonlinear coupling of each particle to it.

\vskip 3pt
We first replace each particle by an auxiliary linear source and eliminate the fields. The resulting stationary energy is a double power series in the sources, with connected correlators as coefficients. An exact source-to-mass transformation then reconstructs the physical potential without further spatial integration. To evaluate the correlators, we exploit stationarity: the exact field of one body fixes the energy linear in the second source, and the linear field response fixes the quadratic energy. Together with the source-to-mass transformation, these terms determine the probe potential and its first backreaction to all orders in $G$. A few higher field responses supply the remaining correlators through $G^{12}$.

\vskip 3pt
The sections is organized as follows. We define the static potential and its gauge convention, introduce the correlators and the source-to-mass transformation, and establish the stationarity and regularization identities used to evaluate the energy. The expansion about one body then provides the all-order first-backreaction proof. Three complementary routes evaluate and cross-check the correlators: route~I uses the exact field of one body, route~II uses two-body fields, and route~III integrates the field pairings analytically. We distinguish analytic results from numerical identifications before converting to the symmetric mass variables of Figure~\ref{fig:mutual} and listing the complete potential in connected correlators. Further details will be presented separately~\cite{ZCXB:inPreparation}.

\subsection*{The object computed}
We compute the static sector of the conservative interaction of two minimally coupled, non-spinning point particles, labelled $A=1,2$, with masses $m_A$ and fixed positions $\mathbf x_A$ separated by a coordinate distance $r$. We use the harmonic-gauge Kaluza-Klein variables \cite{Brunello:2025gpf,Brunello:2026anu}, set $c=1$ except where stated, and work in $d=3+\epsilon$ spatial dimensions, with the $d$-dimensional gravitational constant $G_d$. With $g_{00}=e^{2\phi}$ and $g_{ij}=-e^{(2-c_d)\phi}\gamma_{ij}$\cite{KolSmolkin:2008}, where $c_d=2(d-1)/(d-2)$ and $\gamma_{ij}=\delta_{ij}+\sigma_{ij}$, the static action is
\begin{equation}
 S=-\frac{1}{16\pi G_d}\int\dd t\,\dd^dx\,\sqrt\gamma\Bigl[\frac{c_d}{2}\gamma^{ij}\partial_i\phi\,\partial_j\phi-R[\gamma]+\frac12\gamma_{ij}\Gamma^i\Gamma^j\Bigr]
 -\sum_{A=1,2}m_A\int\dd t\,e^{\phi(\mathbf x_A)},\qquad \Gamma^i=\gamma^{jk}\Gamma^i{}_{jk},
 \label{meth:action}
\end{equation}
with the contracted connection taken in the original Cartesian chart. For static fields the energy functional is $\mathcal E=-S/\Delta t$; subtracting the rest-mass and scaleless self-energy contributions from its stationary value gives the potential $V$. Three properties organize the calculation: the gauge-fixing term contains no $\phi$, because $\sqrt{-g}g^{ij}=-\sqrt\gamma\gamma^{ij}$; the scalar enters the bulk quadratically, so the bulk vertices are $\phi\phi\sigma^k$ ($k\ge1$) or $\sigma^k$ ($k\ge3$), each with two derivatives; and $\sigma$ does not couple to static worldlines.

\vskip 3pt
The fixed positions need not satisfy the particle equations of motion: we extract the velocity-independent sector of the particle effective action off shell. The stresses required to hold the particles in place are not included, so the point-particle stress-energy is not covariantly conserved. We retain the gauge-fixing term and its source contributions and perform no trajectory redefinition. The resulting coefficients define a coordinate-dependent potential rather than a gauge-invariant binding energy. Through $G^8$ they agree with the published static coefficients in the same convention~\cite{GilmoreRoss:2008,FoffaSturani:2011,Foffa:2016rgu,DamourJaranowski:2017,Foffa:2019hrb,Blumlein:2019zku,Brunello:2025gpf,Brunello:2026anu}.

\vskip 3pt
To track the gauge-fixing contribution, we multiply it by $1/\lambda$ for nonzero $\lambda$. The value $\lambda=1$ recovers the action above and is the convention used for the quoted results unless stated otherwise. The constrained case $\lambda=0$ is implemented separately by a Lagrange multiplier that imposes the harmonic condition before variation. At $\lambda=1$, the non-conserved point-particle stress-energy produces a violation of that condition proportional to the forces on the particles. For moving bodies the corresponding dynamical term is quadratic in the equations of motion, affects no gauge-invariant observable and can be removed by redefining the positions~\cite{DamourSchafer:1991}. Thus $\lambda=0$ is one convention among others, not a preferred one, and the static sector remains off shell for every $\lambda$.

\subsection*{From correlators to the potential}
In these field variables, a static particle couples through $m_Ae^{\phi(\mathbf x_A)}$, its mass multiplied by its local proper-time rate relative to coordinate time. Auxiliary linear sources separate this coupling from the bulk-field calculation; the physical coupling is restored exactly afterwards.

\vskip 3pt
\paragraph{Linear sources and correlators} Replacing $m_Ae^{\phi(\mathbf x_A)}$ by a linear source $J_A\phi(\mathbf x_A)$ and eliminating the bulk fields gives the linear-source energy $W$, a double power series in the sources. With the source couplings $x=GJ_1/r$ and $y=GJ_2/r$,
\begin{equation}
 W=-\frac rG\,\Wc(x,y),\qquad \Wc(x,y)=\sum_{a,b\ge1}\frac{\Gb_{a,b}}{a!\,b!}\,x^ay^b ,
 \label{meth:primes}
\end{equation}
normalized by $W=-GJ_1J_2/r+\dots$, so that $\Gb_{1,1}=1$. The coefficient $\Gb_{a,b}$ collects all connected contributions with $a$ insertions of the scalar on body~1 and $b$ on body~2, symmetry weights included; these are the complete connected correlators or primes \cite{Brunello:2026anu}, and $N=a+b$ is their total source degree. The dimensional correlators
\begin{equation}
 \Gamma_{a,b}(r)=\Bigl(\frac Gr\Bigr)^{a+b-1}\Gb_{a,b},\qquad W=-\sum_{a,b\ge1}\frac{\Gamma_{a,b}(r)}{a!\,b!}\,J_1^aJ_2^b ,
 \label{meth:gammadefinition}
\end{equation}
make the leading term $-\Gamma_{1,1}J_1J_2$, with $\Gamma_{1,1}=G/r$; a correlator of degree $N$ first contributes at $G^{N-1}$. These correlators are unrelated to the contracted connection $\Gamma^i$. Exchange symmetry gives $\Gb_{a,b}=\Gb_{b,a}$. The bulk action is invariant under $\phi\to-\phi$\cite{Buchdahl:1959}, so reversing $\phi$ and both linear sources leaves $W$ unchanged and removes odd total source degrees. There are therefore 21 independent correlators through $G^{12}$, labelled with $a\le b$. The exact one-body field, Schwarzschild in the stated convention, fixes the terms of $\Wc$ linear in either source; subtracting $xy$ counts the Newtonian exchange once:
\begin{equation}
 \Wc\supset x\artanh y+y\artanh x-xy,\qquad \Gb_{1,2k+1}=(2k)!.
 \label{meth:probe}
\end{equation}

\vskip 3pt
\paragraph{Restoring the particle couplings} Substituting $m_Ae^\phi=\ext_J\{J\phi-J\log(J/m_A)+J\}$ before the bulk fields are eliminated restores the couplings exactly:
\begin{equation}
 V=\ext_{J_1,J_2}\Bigl\{W-\sum_A\bigl[J_A\log(J_A/m_A)-J_A+m_A\bigr]\Bigr\}.
 \label{meth:sourcefunctional}
\end{equation}
With the physical couplings $\alpha=Gm_1/r$ and $\beta=Gm_2/r$ and with $v=GV/r$, the stationarity conditions, the value and its derivatives are
\begin{equation}
 \alpha=xe^{\Wc_x},\qquad \beta=ye^{\Wc_y},\qquad v=-\Wc+x\Wc_x+y\Wc_y+x+y-\alpha-\beta,\qquad
 \partial_\alpha v=\frac x\alpha-1,\quad \partial_\beta v=\frac y\beta-1 .
 \label{meth:sewing}
\end{equation}
These relations are the classical limit of the correlator formula \cite{Brunello:2026anu}. The map $(x,y)\mapsto(\alpha,\beta)$ has unit Jacobian at the origin and therefore a unique formal inverse. At any fixed perturbative order, reconstruction requires finitely many algebraic operations and introduces no dimensional poles. Only the finite values of the required correlators are needed, not their dimensional continuations. In terms of the dimensional correlators,
\begin{equation}
 V=J_1+J_2-m_1-m_2+\sum_{a,b\ge1}\frac{(a+b-1)\,\Gamma_{a,b}(r)}{a!\,b!}\,J_1^aJ_2^b
 =\sum_{a,b\ge1}c_{a,b}\,\frac{G^{a+b-1}}{r^{a+b-1}}\,m_1^am_2^b .
 \label{meth:potentialGamma}
\end{equation}
Each coefficient $c_{a,b}$ is a polynomial in the correlators: it contains $\Gb_{a,b}$ with coefficient $-1/(a!\,b!)$, and products of lower correlators generated by the exponential couplings.

\vskip 3pt
\paragraph{Coefficient formula} The coefficients follow without solving for the sources. Differentiating equation~\eqref{meth:sewing} gives $\alpha\beta\,\partial_\alpha\partial_\beta v=-xy\Wc_{xy}/\mathcal J$, with $\mathcal J=(1+x\Wc_{xx})(1+y\Wc_{yy})-xy\,\Wc_{xy}^{\,2}$, and, with the coefficient extraction written as a formal residue in $\alpha$ and $\beta$, the change of variables $\dd\alpha\,\dd\beta=e^{\Wc_x+\Wc_y}\mathcal J\,\dd x\,\dd y$ cancels the determinant\cite{Good:1960}, so that
\begin{equation}
 c_{a,b}=-\frac1{ab}\,[x^{a-1}y^{b-1}]\;\Wc_{xy}\,e^{-a\Wc_x-b\Wc_y}.
 \label{meth:coefficient}
\end{equation}
This formula, its Lagrange-inversion form and a direct iteration of the stationarity conditions, implemented independently in exact arithmetic, agree on all 78 mass monomials through $G^{12}$.

\vskip 3pt
\paragraph{The complete reconstruction} If each $\Gamma_{i,j}(r)$ is given the weight $i+j-1$, every monomial of $c_{a,b}$ has weight $a+b-1$, so evaluating $c_{a,b}$ on the dimensional correlators absorbs the factor $(G/r)^{a+b-1}$ in equation~\eqref{meth:potentialGamma}, and the potential at order $G^n$ becomes a polynomial in the masses and the correlators,
\begin{equation}
 V=\sum_{n=1}^{12}V_{G^n}+O(G^{13}),\qquad V_{G^n}=\sum_{a=1}^{n}c_{a,n+1-a}[\Gamma(r)]\,m_1^am_2^{n+1-a},
 \label{meth:gammaexpansion}
\end{equation}
with $c_{a,b}[\Gamma(r)]$ denoting $c_{a,b}$ evaluated on the $\Gamma_{i,j}(r)$. Because correlators of odd degree vanish, the step from $G^{2p-1}$ to $G^{2p}$ introduces no new correlator, so the correlators through degree twelve determine the potential through $G^{12}$. The last section lists $V_{G^1}$ to $V_{G^{12}}$ in full.

\subsection*{Stationarity and field pairings}
Because the first variation of the energy vanishes at the solution, an energy coefficient is fixed before the corresponding field expansion is complete. This section uses that property to replace the sum over Feynman diagrams by a few pairings of field responses.

\vskip 3pt
\paragraph{Diagrams} With linear sources, the correlators are sums of connected tree diagrams of the bulk fields with $N$ source endpoints, which become $(N-2)$-loop integrals in momentum space. Two independent enumerators, which reproduce every published count and satisfy a sum rule over automorphism orders in every sector at $G^7$ and $G^9$, give 282, 4,235 and 74,092 prime diagrams modulo exchange at $G^7$, $G^9$ and $G^{11}$, counted before scaleless terms are discarded. At $G^9$ the standard reductions leave seven- and eight-loop cores in the sectors $(4,6)$ and $(5,5)$, which the reorganization below avoids.

\vskip 3pt
\paragraph{Field form} The sum of all prime diagrams of degree $N$ is the degree-$N$ part $W^{(N)}$ of the stationary value $W[J]$ of the linear-source energy\cite{BoulwareBrown:1968}, with source density $J(\mathbf x)=\sum_AJ_A\delta^{(d)}(\mathbf x-\mathbf x_A)$. Because $\delta W/\delta J=\phi_\star$ at the solution, $W^{(N)}=N^{-1}\int J\,\phi_{(N-1)}$, where the solution is expanded in total source degree, $\phi_\star=\phi_{(1)}+\phi_{(3)}+\phi_{(5)}+\cdots$ and $\sigma_\star=\sigma_{(2)}+\sigma_{(4)}+\sigma_{(6)}+\cdots$. The bulk energy functional is
\begin{equation}
 \mathcal E_{\rm bulk}=\tfrac12\langle\phi,K_\phi\phi\rangle+\tfrac12\langle\sigma,K_\sigma\sigma\rangle+\sum_{n\ge3}\frac1{n!}\,\mathcal T_n[\sigma^n]+\sum_{k\ge1}\frac1{2\,k!}\,\mathcal C_k[\sigma^k;\phi,\phi],
 \label{meth:bulkpairings}
\end{equation}
with kinetic operators $K_\phi$ and $K_\sigma$ and symmetric multilinear forms $\mathcal T_n$ and $\mathcal C_k$ of the tensor and mixed vertices. Cutting a diagram at a propagator turns either side into a rooted branch, which is a field response: the vertices build its source, and the propagator inverts the kinetic operator. Collecting equal branches into $\phi_{(2k+1)}$ and $\sigma_{(2k)}$ turns the diagram sum into pairings of these responses.

\vskip 3pt
\paragraph{The $2n+1$ principle} Let $e$ collect the scalar and tensor field perturbations. Because the energy is stationary in the fields, a field error $\delta e$ of order $p+1$ in the sources changes it by $\frac12\mathcal E''[\delta e,\delta e]+O(\delta e^3)$, of order $2p+2$, so field responses through order $p$ fix the energy through order $2p+1$\cite{Wigner:1935,DalgarnoStewart:1956,Gonze:1995variational}. The argument needs fixed boundary and contact data, not positivity of the Hessian, and the principle is applied before the finite part is taken. It holds in both expansions of the energy, in total source degree and about the exact field of one body.

\vskip 3pt
\paragraph{Stationarity reduction} By the $2n+1$ principle, $W^{(N)}$ is unchanged when the fields are truncated at degree $N/2$, and the highest of them then enters, once its field equation is used, only through one quadratic form. At $G^7$ ($N=8$) the terms of the truncated energy that contain $\sigma_{(4)}$ are $\frac12\langle\sigma_{(4)},K_\sigma\sigma_{(4)}\rangle+\frac12\mathcal T_3[\sigma_{(2)},\sigma_{(2)},\sigma_{(4)}]+\mathcal C_1[\sigma_{(4)};\phi_{(1)},\phi_{(3)}]+\frac12\mathcal C_2[\sigma_{(2)},\sigma_{(4)};\phi_{(1)},\phi_{(1)}]$; being stationary in $\sigma_{(4)}$, they equal $-\frac12\langle\sigma_{(4)},K_\sigma\sigma_{(4)}\rangle$. Thus
\begin{equation}
 W^{(8)}=-\tfrac12\langle\sigma_{(4)},K_\sigma\sigma_{(4)}\rangle+\tfrac1{24}\mathcal T_4[\sigma_{(2)}^4]+\tfrac12\mathcal C_1[\sigma_{(2)};\phi_{(3)},\phi_{(3)}]+\tfrac12\mathcal C_2[\sigma_{(2)},\sigma_{(2)};\phi_{(1)},\phi_{(3)}]+\tfrac1{12}\mathcal C_3[\sigma_{(2)}^3;\phi_{(1)},\phi_{(1)}],
 \label{meth:pairing}
\end{equation}
and in the same way $W^{(10)}$ consists of 11 pairings, with $\phi_{(5)}$ entering through $-\frac12\langle\phi_{(5)},K_\phi\phi_{(5)}\rangle$, and $W^{(12)}$ of 18, with $\sigma_{(6)}$ entering through $-\frac12\langle\sigma_{(6)},K_\sigma\sigma_{(6)}\rangle$. The fields required are $\phi_{(1)}$, $\sigma_{(2)}$, $\phi_{(3)}$ and $\sigma_{(4)}$ at $G^7$, together with $\phi_{(5)}$ at $G^9$ and $\sigma_{(6)}$ at $G^{11}$.

\begin{figure}[t]
 \centering
 \includegraphics[width=0.694\textwidth]{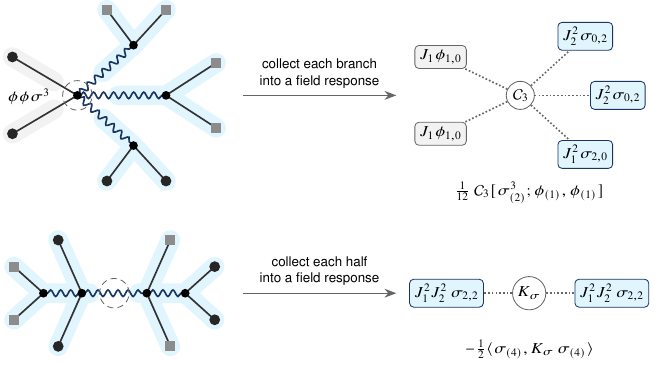}
 \caption{\textbf{Feynman diagrams become pairings of field responses.} \textit{Top}. A diagram at $G^7$ with four insertions on each body (circles, body~1; squares, body~2; solid lines, scalar $\phi$; wavy lines, metric $\sigma$). Cut at its centre, a $\phi\phi\sigma^3$ vertex (dashed circle), it falls into branches, each shaded like the field response it becomes (\textit{right}): the three metric branches are the components $\sigma_{0,2}$, $\sigma_{0,2}$ and $\sigma_{2,0}$ of the metric response $\sigma_{(2)}$, of degree two in the sources, and the two direct legs are the Newtonian field $\phi_{1,0}$ of body~1. The diagram is thereby one term of the pairing $\frac1{12}\mathcal C_3[\sigma_{(2)}^3;\phi_{(1)},\phi_{(1)}]$ of equation~\eqref{meth:pairing}: the pairing contributes $\frac1{12}\times\frac{3!}{2!}=\frac14$, and each of the three branches with two equal sources a factor $\frac12$, which gives the symmetry weight $1/32$ of the diagram. \textit{Bottom}. A diagram of the same sector centred on a metric propagator (dashed circle). Its two halves are different terms of the component $\sigma_{2,2}$ of $\sigma_{(4)}$, so the diagram is one term of the first pairing of equation~\eqref{meth:pairing}, $-\frac12\langle\sigma_{(4)},K_\sigma\sigma_{(4)}\rangle$, which contributes $\frac12\times2=1$; each half contributes a factor $\frac14$ from its two pairs of equal sources, which gives the symmetry weight $1/16$.}
 \label{fig:pairing}
\end{figure}

\vskip 3pt
\paragraph{Diagram by diagram} Each diagram is assigned to exactly one pairing by its centre: either the propagator that divides its source endpoints into equal halves, placing it in the highest quadratic pairing, or the vertex from which every branch carries fewer than half of them. Expanding the pairing in the sources reproduces the diagram with its symmetry weight (Figure~\ref{fig:pairing}). Evaluation then proceeds by source sector. With $\phi_{(n)}=\sum_{a+b=n}J_1^aJ_2^b\phi_{a,b}$ and the analogous expansion for $\sigma_{(n)}$, equation~\eqref{meth:gammadefinition} gives $[J_1^aJ_2^b]W^{(N)}=-\Gamma_{a,b}(r)/(a!\,b!)$ for $N=a+b$, which fixes $\Gb_{a,b}$ with its normalization. Each field response is solved once and reused in every pairing and sector in which it occurs.

\vskip 3pt
\paragraph{Pairing identities} Fields that enter only through pairings need not be solved for: for a formally self-adjoint operator $\mathscr D=-\partial_iA^{ij}\partial_j+\mathcal U$ with $\mathscr Df=J$ and $\mathscr Dg=w$, on a domain $D_\delta$ punctured by true spheres $|\mathbf x-\mathbf x_A|=\delta$ around the sources,
\begin{equation}
 \int_{D_\delta}wf=\int_{D_\delta}gJ+\int_{\partial D_\delta}\bigl(gA^{ij}\partial_jf-fA^{ij}\partial_jg\bigr)n_i\,\dd S ,
 \label{meth:adjoint}
\end{equation}
so formal self-adjointness does not justify dropping the surface terms. Route~III uses these identities to replace a required response integral by a pairing with the dual solution $g$, when the latter is simpler to obtain.

\subsection*{Dimensional regularization}
Point-source singularities leave finite contributions as well as poles, and both must be kept. All evaluations compute the dimensionally regulated value\cite{tHooftVeltman:1972} of each correlator, in which the poles of individual terms cancel, and realize it as a three-dimensional finite part of its integrand $I$ plus local corrections $\Delta_A$ at the worldlines,
\begin{equation}
 W^{\DR}=\Pf\!\int\!\dd^3x\,I+\sum_A\Delta_A ,
 \label{meth:dr}
\end{equation}
with Riesz powers\cite{Riesz:1949} continued jointly from a common convergence domain, the dimension kept in every normalization, contraction and angular average, and true spheres around the sources as reference excisions. The split introduces a length scale at each worldline, the subtraction scale, on which the two parts depend separately but their sum does not. Boundary fluxes of integrations by parts and contact terms of distributional derivatives stay with the pairing in which they arise. Four points govern the local corrections.

\vskip 3pt
\emph{Poles are local and cancel.} Terms with an odd number $j\ge3$ of singular self-field factors (legs) of one body produce poles there, proportional to $1/[(1-j)\epsilon]$, which cancel as polynomials in the regular data of the other fields, without counterterms.

\vskip 3pt
\emph{$\epsilon\times\epsilon^{-1}$ remnants survive.} After the cancellation, finite terms remain that depend on the $O(\epsilon)$ parts of the singular local data. For a logarithmic radial layer at distance $\rho$ from a source, with $\kappa\ne0$ ($\kappa=1-j$ for the critical densities),
\begin{gather}
 \int_0^b(a_0+\epsilon a_1+\cdots)\rho^{-1+\kappa\epsilon}\,\dd\rho=\frac{a_0}{\kappa\epsilon}+a_0\log b+\frac{a_1}\kappa+O(\epsilon),\notag\\
 \FP_{\epsilon=0}\frac{(a_{-1}/\epsilon+a_0+\epsilon a_1)\,b^{\kappa\epsilon}}{\kappa\epsilon}=\frac{a_1}\kappa+a_0\log b+\frac{\kappa a_{-1}}2\log^2b .
 \label{meth:radialfinite}
\end{gather}
Integrands must therefore not be set to $d=3$ first, and a purely three-dimensional Hadamard computation misses these finite terms once a point singularity carries $j\ge3$ legs; in the nested-pole case the poles cancel but $a_1$ remains. In general, a regular factor that multiplies a pole of order $p$ must be kept through $O(\epsilon^p)$; the factor $1/(d-2)=1-\epsilon+\cdots$ of the tensor propagator, for instance, shifts the finite part of a simple pole $b/\epsilon$ by $-b$. Mode solutions must likewise be built in $d=3+\epsilon$: three-dimensional multipole solutions give $\Gb_{2,2}=28/3$ instead of 8.

\vskip 3pt
\emph{The split depends on the representation.} Integration by parts and changes of variables or excision geometry alter the finite part, which must therefore be combined with the local correction in the same representation. The two contributions may separately contain transcendental constants. For the $(5,5)$ sector, with both subtraction scales set to $r$ and in the normalization $\Gb_{5,5}/(5!)^2$, the three-dimensional finite part and local correction are respectively $\frac{5419906609}{3472875}-\frac{499\pi^2}{1890}$ and $\frac{953976641}{3472875}+\frac{499\pi^2}{1890}$. Their sum is $\frac{5506}3$: the $\pi^2$ terms cancel only after the local correction is included.

\vskip 3pt
\emph{Double poles require global regular data.} In sectors with many source insertions on one worldline, the critical angular averages can themselves carry poles. The total pole still cancels, but the finite part can then depend on the $O(\epsilon)$, and where poles are nested the $O(\epsilon^2)$, parts of the regular data of the two-body fields at that point, which local expansions do not determine: the regular homogeneous parts of a field come from its global solution. This regime first occurs at $G^{11}$, in the sectors of $\Gb_{3,9}$, $\Gb_{4,8}$ and $\Gb_{5,7}$, at the body that carries nine, eight and seven insertions; in $\Gb_{5,7}$ the dependence cancels in the complete local sum. Routes~I and~II locate it with a sensitivity check: they perturb the regular data of every two-body field by $\epsilon$ times generic harmonics and record which perturbations change the result. Only perturbations of the metric field bilinear in the two sources do. The check only locates the dependence; it is fixed by the exact $d$-dimensional regular data of this field, which follow in closed form from the hard region of its momentum representation. Without them, two numerical evaluations of $\Gb_{3,9}$ differ by exactly 4096. Route~III keeps unknown regular constants symbolic and shows that they drop out, by angular parity, wherever the single-pole regime holds; otherwise it takes the required regular coefficients from the global solutions, to the depth that the poles set.

\subsection*{Expansion about one body}
The proof of the first backreaction and route~I expand the energy about the exact field of one body. At $d=3$ the field of body~1 is harmonic-gauge Schwarzschild\cite{Fock:1959}; field-theory recursions reconstruct it from its weak-field expansion\cite{Duff:1973,DamgaardLee:2024,MougiakakosVanhove:2024}. With the background coupling $\mathfrak m$, equal to $GJ_1$ at $d=3$, and the harmonic radius $\rho$ about body~1,
\begin{equation}
 \phi_0=-\artanh\frac{\mathfrak m}\rho,\qquad \gamma_0=\dd\rho^2+(\rho^2-\mathfrak m^2)\,\dd\Omega^2,\qquad
 \sqrt\gamma\,\gamma^{ij}\big|_0=\delta^{ij}-\frac{\mathfrak m^2}{\rho^2}\,n^in^j ,
 \label{meth:background}
\end{equation}
with $n^i$ the radial unit vector. In the gothic variables $\mathfrak p^{ij}=\sqrt\gamma\gamma^{ij}=\delta^{ij}-h^{ij}$ the background is exactly quadratic in $\mathfrak m$, the scalar couples to $h$ through the single vertex $(\delta-h)^{ij}\partial_i\phi\,\partial_j\phi$, and the harmonic condition is linear. In Weyl canonical coordinates\cite{Weyl:1917}, $\varrho=\sqrt{\rho^2-\mathfrak m^2}\sin\theta$ and $Z=\rho\cos\theta$, the background scalar operator on axisymmetric functions is the flat Laplacian, and the scalar Green function is elementary\cite{Wiseman:2000} (electrostatic analogue in~\cite{Copson:1928,Linet:1976}), $\mathscr G(\mathbf x,\mathbf x')=\{4\pi[|\mathbf x-\mathbf x'|^2-\mathfrak m^2|\mathbf x\times\mathbf x'|^2/(\rho^2\rho'^2)]^{1/2}\}^{-1}$. Expanded in $j=J_2$ about this background (energy $\mathcal E_0$), with scalar and tensor perturbations $e$, the potential functional is
\begin{equation}
 \mathcal E(e;j)=\mathcal E_0+j\phi_0(\mathbf x_2)+j\,s(e)+\tfrac12\langle e,\calH e\rangle+\sum_{k\ge3}T_k(e),
 \label{meth:functional}
\end{equation}
where $s(e)$ is the scalar perturbation at source~2, $\calH$ is the gauge-fixed Hessian, whose Green inverse is selected by dimensional continuation, and $T_k$ is the Taylor term of order $k$, including $1/k!$. The stationary point $e(j)=\sum_kj^ke_k$ defines the field responses, which obey $\calH e_1=-s$, $\calH e_2=-T_3'(e_1)$ and $\calH e_3=-\mathscr S_3$, with $\mathscr S_3=T_4'(e_1)+T_3''(e_1)e_2$, and the coefficients $W_k=[j^k]W$ are
\begin{align}
 W_2&=-\tfrac12\langle s,\calH^{-1}s\rangle,\qquad W_3=T_3(e_1),\qquad
 W_4=T_4(e_1)+\tfrac12T_3'(e_1)e_2,\nonumber\\
 W_5&=T_5(e_1)+T_4'(e_1)e_2+\tfrac12T_3''(e_1)[e_2,e_2],\qquad
 W_6=T_6(e_1)+T_5'(e_1)e_2+\tfrac12T_4''(e_1)[e_2,e_2]+T_3(e_2)+\tfrac12\langle \mathscr S_3,e_3\rangle ,
 \label{meth:hierarchy}
\end{align}
from which the correlators follow through $W_b(J_1)=-\sum_a\Gb_{a,b}G^{a+b-1}J_1^a/(a!\,b!\,r^{a+b-1})$. Every $G^9$ correlator therefore needs at most the second field response. Among the correlators required through $G^{12}$, only $\Gb_{6,6}$ needs the third.

\subsection*{The all-order first backreaction}
The energy quadratic in the second source supplies the field contribution needed for the first backreaction. It is evaluated on the exact background of the first source and requires only the linear field response, evaluated at the second source. Restoring the particle couplings then gives the physical 1SF potential. We first establish the quadratic energy:
\begin{equation}
 W\big|_{J_2^2}=-\frac{GJ_2^2}{2r}f_2(x),\qquad
 f_2(x)=\sum_{k\ge1}\Gb_{2,2k}\frac{x^{2k}}{(2k)!}=\frac{4x^2}{(1-x^2)^2}\Bigl(1-\frac x3\artanh x\Bigr),
 \label{eq:f2}
\end{equation}
whose expansion reproduces $\Gb_{2,2}=8$, $\Gb_{2,4}=160$ and $\Gb_{2,6}=6400$ and gives the entire tower: $\Gb_{2,8}=437248$, $\Gb_{2,10}=45757440$ and, in general, $\Gb_{2,2k}/(2k)!=\frac23[(7k-1)-(2k-1)\sum_{j=1}^{k-1}(2j-1)^{-1}]$.

\vskip 3pt
We establish equation~\eqref{eq:f2} coefficient by coefficient within dimensionally regularized perturbation theory. The proof uses the splitting of singular regions with meromorphic continuation in the dimension, the vanishing of scaleless integrals, uniform bounds on the subtracted angular sums, joint holomorphy in the multipole degree $\ell$ and in $n=d-2$, and exact algebraic identities of the coupled radial equations.

\vskip 3pt
The five steps below first construct the regulated angular-mode sum and justify continuation in $\ell$. The subtracted generic modes, continued to all $\ell\ge0$, then sum to zero. The regulated result is therefore determined by the differences between the true and continued monopole and dipole contributions; at $\lambda=0$ these differences cancel as well. A Ward identity independently supplies the gauge-fixing contribution. Finally, the source-to-mass transformation converts the quadratic energy into equation~\eqref{eq:physical}.

\paragraph{Steps 1 and 2: regulated response and angular sum} The coupled scalar--tensor Green function $\mathcal G=(16\pi G_d\calH)^{-1}$ follows by Born recursion in $\mathfrak m$ and is selected by dimensional continuation of the flat inner and outer families, order by order in the background, so that no separate boundary condition at the horizon enters. The quadratic energy is determined by the regulated coincidence value of the scalar entry of $\mathcal G$, which therefore includes the metric response: $W|_{J_2^2}=-GJ_2^2f_2^{(\lambda)}(x)/(2r)=-8\pi G_dJ_2^2[\mathcal G^{\phi\phi}(\mathbf x_2,\mathbf x_2)]_{\DR}$. Because the full homogeneous convolution at the last vertex is scaleless, and so vanishes in dimensional regularization, this coincidence value is the regular part of the point-split field. Projection onto angular modes then gives
\begin{equation}
 [x^k]f_2^{(\lambda)}=\lim_{n\to1}\sum_{\ell\ge0}\frac{(n)_\ell}{\ell!}[\mathcal F_\ell^{(k)}(n)-\beta_{{\rm ang},k}(n)]=\sum_{\ell\ge0}[\mathcal F^{(k)}_{{\rm true},\ell}-b_k],\qquad \sum_kb_kx^k=(1-x^2)^{-1/2},
 \label{meth:angular}
\end{equation}
where $(n)_\ell/\ell!=C^{(n/2)}_\ell(1)$, with $(n)_\ell=n(n+1)\cdots(n+\ell-1)$, is the Gegenbauer weight of degree $\ell$, and $\mathcal F_\ell^{(k)}$ is the coincident radial coefficient divided by its flat Green factor, with limit $\mathcal F^{(k)}_{{\rm true},\ell}$ as $n\to1$ at fixed $\ell$. The subtraction $\beta_{\rm ang}=(PQ)^{-1/2}$, with $\mathfrak p^{\rho\rho}=\rho^{n+1}P$ and $\mathfrak p^{\theta\theta}=\rho^{n-1}Q$ for the background, is fixed by the local singularity of the Green function. Summed over modes, it has no constant term at coincidence for non-integer $n$ and so leaves the regulated value unchanged, like the regularization parameters of mode-sum regularization\cite{BarackOri:2000,PoissonPoundVega:2011}.

\vskip 3pt
\paragraph{Step 3: continuation in $\ell$} Taking the limit $n\to1$ inside the sum, the second equality in equation~\eqref{meth:angular}, requires two facts. First, the tail is uniform: at fixed even $k$, the Riccati recursion for the radial logarithmic derivatives gives $\mathcal F_\ell^{(k)}-\beta_{{\rm ang},k}=L^{-2}h_k(L^{-2},n)$, even and rational in $L=2\ell+n$ and bounded for large $|L|$ near $n=1$, so the tail, $O(\ell^{{\rm Re}\,n-3})$, is summable. The bound is obtained in the non-singular gauges $\lambda=\frac12$ and $1$ and extends to $\lambda=0$ because each subtracted coefficient is affine in $\lambda$. Second, for $\ell\ge2$ the fixed-mode limits coincide with the continuation in $\ell$. The two could differ only at resonances, where continued inner and outer radial solutions collide, which the flat indicial roots confine to the background orders $2\ell-3\le k\le2\ell+5$. A pole within one solution space is removed by a change of basis, and the conserved radial pairing, the boundary form of the radial Hessian, excludes poles into the opposite space. For the one scalar family at risk, the static tidal solution, every possible obstruction is proportional to the trilinear residue
\begin{equation}
 \Res_{z=\infty}\bigl[(z^2-1)P_m'(z)\,\mathcal P_\sigma'(z)\,\mathcal P_{m-\sigma}'(z)\,\dd z\bigr]=0,
 \label{meth:trilinear}
\end{equation}
in the variable $z=2(\mathcal R/\mathcal R_s)^n-1$, with $\mathcal R$ the areal radius and $\mathcal R_s^n=2\mathfrak m$, where $P_m$ is the Legendre polynomial and $\mathcal P_\sigma$ the monic growing Legendre series of degree $\sigma$; it vanishes identically, by a Legendre identity.

\vskip 3pt
\paragraph{Step 4: the continued modes sum to zero} The generic mode, continued in $\ell$, is $\mathcal F_{\rm gen}=\sum_{i=0}^2(\gamma_i\partial_z^i\Pi+\eta_i\partial_z^i\partial_\ell\Pi)$, with $\Pi=\widetilde P_\ell Q_\ell$, $\widetilde P_\ell=-\pi^{-1}\tan(\pi\ell)Q_{-\ell-1}$, $Q_\ell$ the Legendre function of the second kind, $z=1/x$, and rational coefficients $\gamma_i,\eta_i$ obtained by pairing the exact inner and outer families. Each $\widehat R_k(L)=[x^k]\mathcal F_{\rm gen}-b_k$ is even, rational and $O(L^{-2})$, with only even-integer poles of order at most three; the individual coefficient functions also have poles at $L=1$, which cancel in the complete generic mode. Integrating $\widehat R_k(L)\frac\pi2\tan\frac{\pi L}2$ around expanding circles gives
\begin{equation}
 \sum_{\ell\ge0}\widehat R_k(2\ell+1)=\frac{\pi^2}8\mathsf A_{2,k},\qquad \mathsf A_{2,k}=\sum_p[(L-p)^{-2}]\widehat R_k .
 \label{meth:contour}
\end{equation}
The continued modes therefore sum to zero if $\mathsf A_{2,k}$ vanishes, a functional identity of the radial pairing with a finite certificate. The residue theorem, applied to the terms $\eta_i\partial_z^i\Pi$, transfers the coupling-dependent poles of $\Pi$ to the fixed poles of $\gamma_i$ and $\eta_i$, at $L=1$, 2 and 4, whose residues are rational functions of $z$ proportional to $c_\lambda=(1+\lambda)/2$. Using $\partial_\ell=2\partial_L$ and collecting all orders, $\mathsf A_2=\sum_k\mathsf A_{2,k}x^k$ becomes
\begin{equation}
 \mathsf A_2=2\sum_{p=2,4}\sum_{i=0}^2\bigl[(\Res_p\gamma_i)\,\partial_z^i\Res_p\Pi+2(\Res_p\eta_i)\,\partial_z^i\FP_p\Pi\bigr]+4\sum_{i=0}^2(\Res_1\eta_i)\,\partial_z^iQ_0 ,
 \label{meth:localized}
\end{equation}
where the residue $\Res_p\Pi$ and the constant Laurent coefficient $\FP_p\Pi$ of $\Pi$ at $L=p$ are quadratic in $\mathcal A=Q_{-1/2}(z)$ and $\mathcal B=Q_{1/2}(z)$, with coefficients polynomial in $z$, and the $p=1$ term vanishes. The terms from $\gamma_i$ and from $\eta_i$ then give $\frac{\pi^2}4\mathsf A_2=c_\lambda\Theta(z)-c_\lambda\Theta(z)=0$, with $\chi=z^2-1$ and
\begin{equation}
 \Theta(z)=\frac{2z\,[(2z^2+25)\mathcal A^2-2z(8z^2-17)\mathcal A\mathcal B+(32z^4+4z^2-81)\mathcal B^2]}{27\chi^2},
 \label{meth:certificate}
\end{equation}
so that the three rational coefficients, of $\mathcal A^2$, $\mathcal A\mathcal B$ and $\mathcal B^2$, cancel identically. This finite identity is the collective cancellation: at every order in $x$, the subtracted generic modes, continued to every $\ell$, sum to zero.

\vskip 3pt
\paragraph{Step 5: exceptional modes} By steps~3 and~4, only the corrections to the continued monopole and dipole remain: equation~\eqref{meth:angular} reduces to $\Delta_0+\Delta_1$, with $\Delta_\ell=\sum_kx^k[\mathcal F^{(k)}_{{\rm true},\ell}-\mathcal F^{(k)}_{\rm gen}|_{L=2\ell+1}]$. At $\ell=0,1$, where some tensor harmonics vanish identically, generic continuation does not give the regulated source spaces, as for low multipoles in self-force theory\cite{DetweilerPoisson:2004}. Matching the true source spaces gives, with $X=x/(1-x^2)$,
\begin{equation}
 \Delta_0=-4X^2,\qquad \Delta_1=X^2\Bigl[4+\lambda\Bigl(4-\frac43x\artanh x\Bigr)\Bigr],\qquad f_2^{(\lambda)}=\Delta_0+\Delta_1 .
 \label{meth:exceptional}
\end{equation}
For the monopole, at $\lambda=1$, the true and the continued pairings differ only in a numerator term, $3z^2(13z^2-15)$ and $3z^2(13z^2+5)$ respectively, so that $\Delta_0=3z^2(-20)/[15(z^2-1)^2]=-4X^2$. For the dipole at $\lambda=0$, the uniform-field dipole $\delta\phi=c_0Z$, $\delta\mathfrak p^{ij}=0$ ($Z$ along the axis), with multiplier $\Lambda_j=c_dc_0\phi_0\delta_{jZ}$ for the harmonic condition, solves the response equations in every dimension, so $\Delta_1=4X^2=-\Delta_0$ and $f_2^{(0)}=0$: at $\lambda=0$ the quadratic response vanishes, a coupled counterpart of the vanishing self-force on a static scalar charge outside a black hole\cite{Wiseman:2000}. Thus $f_2^{(\lambda)}=\lambda f_2$, and at $\lambda=1$ equation~\eqref{meth:exceptional} gives equation~\eqref{eq:f2}; the Ward identity below gives the same $\lambda$ dependence without the dipole matching at $\lambda\ne0$. At order $x^2$, for example, the true modes sum to $[x^2]f_2=4$ and the continued modes to zero; the two differ only by the monopole and dipole corrections, $-4$ and $+8$, while the true monopole and dipole themselves contribute only $109/105$.

\vskip 3pt
\paragraph{Ward identity: the gauge-fixing contribution} Let $\mathscr W_2(\lambda)=W|_{J_2^2}$ at gauge-fixing weight $\lambda$, let $\mathscr G$ be the Green function of $-\partial_i\mathfrak p^{ij}\partial_j$ on the background, with $\mathscr G_A=\mathscr G(\mathbf x,\mathbf x_A)$, and let $\mathscr F_d=J_2\phi_0'(r;d)$ be the radial derivative of the background interaction energy $J_2\phi_0$ of source~2. The linear Noether identity fixes the gauge violation, $\mathcal V_j=16\pi G_d\lambda\mathscr F_d\delta_{jZ}(\mathscr G_2-\mathscr G_1)$, and differentiation with respect to $\lambda$ gives
\begin{equation}
 \frac{\partial\mathscr W_2}{\partial\lambda}=-8\pi G_d\mathscr F_d^2\,\mathcal I_d,\qquad
 \mathcal I_d=\int_{\DR}\mathfrak p^{ZZ}(\mathscr G_2-\mathscr G_1)^2\,\dd^dx ,
 \label{meth:ward}
\end{equation}
independent of $\lambda$. Hence $\mathscr W_2(1)=\mathscr W_2(0)-8\pi G_d\mathscr F_d^2\mathcal I_d$, where $\mathscr W_2(0)\to0$ as $d\to3$ by steps~4 and~5, and it remains to evaluate $\mathcal I_d$. At $d=3$, $\mathscr G_1=Q_0(\rho/\mathfrak m)/(4\pi\mathfrak m)$ and $\mathfrak p^{ZZ}=1-(\mathfrak m^2/\rho^2)\cos^2\theta$. Subtracting the flat counterparts, with scaleless self terms and the flat mixed integral $-r/(8\pi)$, gives $\mathcal I=\frac r{4\pi}+I_2-2I_1$, where $I_2$ and $I_1$ are the subtracted integrals of $\mathfrak p^{ZZ}\mathscr G_2^2$ and $\mathfrak p^{ZZ}\mathscr G_1\mathscr G_2$. In Weyl coordinates, $\mathfrak p^{ZZ}\,\dd^3x=\varrho\,\dd\varrho\,\dd Z\,\dd\varphi$ and $\mathscr G_2$ is the flat Green function, so $I_2$ maps onto its flat counterpart and vanishes. Only the $\ell=0,2$ projections of $\mathscr G_2$ enter $I_1$; with $Q_\ell=Q_\ell(1/x)$, the Heine expansion\cite{Hobson:1931} gives its three-dimensional finite part, $\frac{7\mathfrak m}{72\pi}(Q_0-Q_2)$.

\vskip 3pt
The local correction to $I_1$ shows the $\epsilon\times\epsilon^{-1}$ mechanism in its simplest form. Writing $\mathfrak p^{ij}=\delta^{ij}-\sum_s(a_s\delta^{ij}+b_sn^in^j)\xi^{2s}$ and $\phi_0=-\xi+p_1\xi^3+\cdots$, with $\xi=\mathfrak m\rho^{-(d-2)}$, the first-order dimensional data of the background are $a_1=-\frac12\epsilon$, $b_1=1+2\epsilon$, $b_2=\frac32\epsilon$ and $p_1=-\frac13-\frac12\epsilon$, and the first quadrupole dressing of the scalar is $e_{\rm Q}=[a_1d-(a_1+b_1)(4-d)]/[(d-2)(6-d)]=-\frac13-\frac49\epsilon$. The critical angular coefficients $Y_0=-p_1-a_1-b_1/d=\frac49\epsilon$ and $Y_2=-p_1b_1+b_2+b_1e_{\rm Q}=\frac{14}9\epsilon$ vanish at $d=3$ but multiply radial poles $-1/(2\epsilon)$ and $-1/(4\epsilon)$; with projections $+Y_0Q_0$ and $-Y_2Q_2$ and normalization $\mathfrak m/(4\pi)$, they give the local correction $\delta I_1=-\frac{\mathfrak m}{18\pi}Q_0+\frac{7\mathfrak m}{72\pi}Q_2$ at body~1. The quadrupole then cancels: $I_1=\frac{\mathfrak m}{24\pi}Q_0=\frac{\mathfrak m}{24\pi}\artanh x$ and $\mathcal I=\frac1{4\pi}(r-\frac{\mathfrak m}3\artanh x)$, so that $\mathscr W_2(1)=-2Gr[J_2\mathfrak m/(r^2-\mathfrak m^2)]^2(1-\frac x3\artanh x)$, which is equation~\eqref{eq:f2}. The uniform $O(\ell^{-2})$ tail of the subtracted generic mode on each disk $|x|\le x_\star<1$ allows the coefficientwise identity to be summed throughout $|x|<1$.

\vskip 3pt
\paragraph{From the quadratic response to the first backreaction} The quadratic linear-source energy supplies the field contribution to the first backreaction. Restoring the particle couplings through equation~\eqref{meth:sewing} adds the source feedback at fixed physical masses, without further integration. With the mass ratio $q=m_2/m_1$, $\beta=q\alpha$, and through quadratic order in $y$, $\Wc=yA(x)+\frac12y^2\lambda f_2(x)$ with $A=\artanh$. Writing $Z_0=e^{-A}$, with all functions evaluated at $\alpha$, the stationarity conditions~\eqref{meth:sewing} give $x=\alpha-\beta\alpha Z_0A'+O(\beta^2)$ and $y=\beta Z_0+\beta^2Z_0^2(\alpha A'^2-\lambda f_2)+O(\beta^3)$. Integrating the envelope identity $\partial_\beta v=y/\beta-1$ at fixed $\alpha$ gives $V/m_2=\Phi_0(\alpha)+q\Phi_1^{(\lambda)}(\alpha)+O(q^2)$, with
\begin{equation}
 \Phi_0=\sqrt{\frac{1-\alpha}{1+\alpha}}-1,\qquad \Phi_1^{(\lambda)}=\frac{\alpha Z_0^2}2(\alpha A'^2-\lambda f_2)=\frac12(\alpha\Phi_0')^2\Bigl[1-\lambda\Bigl(4\alpha-\frac43\alpha^2\artanh\alpha\Bigr)\Bigr],
 \label{meth:fixedmass}
\end{equation}
where $\alpha Z_0A'=-\alpha\Phi_0'$ was used. Here $Z_0=1+\Phi_0$ is the redshift factor of the probe, $Z_0^2$ restores the coupling of the lighter mass, $\alpha A'^2=\alpha/(1-\alpha^2)^2$ records the change of the heavier source at fixed $m_1$ and gives the feedback $\frac12(\alpha\Phi_0')^2$, and $-\lambda f_2$ is the bulk response. At $\lambda=1$, $\Phi_1^{(\lambda)}$ is $\Phi_1$ of equation~\eqref{eq:physical}, whose Taylor coefficients are the coefficients $c_{n-1,2}=c_{2,n-1}$.

\vskip 3pt
\paragraph{Dependence on the gauge-fixing weight} The entire $\lambda$ dependence of $\Phi_1^{(\lambda)}$ is the quadratic-response term, which first enters at $G^3$: $[\alpha^3]\Phi_1^{(\lambda)}=-1-2\lambda$ and $[\alpha^4]\Phi_1^{(\lambda)}=2+4\lambda$. With the harmonic condition imposed exactly, $\Phi_1^{(0)}$ follows from the probe function alone. Route~I's per-mode computation, repeated at other weights, checks the $\lambda$ dependence through $x^{10}$, independently of the closed form.

\subsection*{Evaluation of the correlators}
Three complementary implementations evaluate the correlators. Route~I expands about the exact field of one body, using numerical evaluation and reconstruction in exact arithmetic. Route~II uses the two-body fields, evaluated on a grid and in prolate modes. Route~III evaluates the pairings analytically in both expansions of the stationary energy. A separate momentum-space calculation checks the quadratic sector.

\vskip 3pt
\paragraph{Route I: about the exact one-body field} Route~I evaluates equations~\eqref{meth:background}--\eqref{meth:hierarchy}, constructing the field responses to body~2 as multipole series in $d=3+\epsilon$ and taking $d\to3$ mode by mode. For the quadratic energy, four independent implementations recover $\Gb_{2,2}$ to $\Gb_{2,6}$ without using their known values as inputs and reproduce the higher $\Gb_{2,2k}$ in agreement with equation~\eqref{eq:f2}. The longest calculation sums the same modes as the proof order by order in exact arithmetic through $\Gb_{2,48}$. This correlator first enters at $G^{49}$; together with the known probe tower, source reconstruction also checks the physical 1SF potential at $G^{50}$.

\vskip 3pt
The cubic energy $W_3=T_3(e_1)$ requires only the linear field response. It is evaluated in two ways: with punctures, a Hadamard finite part and local corrections from general-$d$ tensor calculations~\cite{DamourJaranowskiSchafer:2001,BlanchetDamourEspositoFarese:2004}, and sphere by sphere about body~1 in $d$ dimensions. In the latter representation, the angular integral of the cubic density is an exact series in the radius and its logarithm. For $\Gb_{3,K}$ with $K\le9$, this method becomes an exact cubic evaluation: series coefficients computed over finite fields~\cite{FLINT}, recurrences inferred and checked against the computed data, a telescoping certificate~\cite{PetkovsekWilfZeilberger:1996} and rational reconstruction~\cite{WangGuyDavenport:1982,Monagan:2004} reproduce $\Gb_{3,7}$ and $\Gb_{3,9}=1770222336/5$. The recurrence bounds are not proved, and the local correction at body~1 for $K=9$ assumes integral Taylor coefficients in the dimension offset modulo a large prime. Under these assumptions, this evaluation provides an additional check of the analytic determination by route~III.

\vskip 3pt
The quartic to sextic energy terms $W_4$ to $W_6$ are evaluated numerically sphere by sphere. Their per-mode matching constants involve infinite inverse-binomial sums. The local $d$-dimensional contributions are evaluated in multiple precision~\cite{Fousse:2007MPFR}, with exact regular data wherever the sensitivity check identifies dependence on them. Exactly known local structures are subtracted before the remaining tails are fitted.

\vskip 3pt
\paragraph{Route II: the two-body fields} Route~II assembles the terms of $W$ of total source degree 8, 10 or 12 from numerically solved two-body fields: scalar fields of degrees 1, 3 and 5 and metric fields of degrees 2, 4 and 6 in gothic variables. The fields are computed on a prolate spheroidal grid with punctures at the sources~\cite{BrandtBrugmann:1997,AnsorgBrugmannTichy:2004}. A Hadamard finite part defined by spherical excisions~\cite{Hadamard:1923} is combined with general-$d$ local corrections at both worldlines. Each evaluation yields every mass sector at the chosen source degree, so known sectors provide checks.

\vskip 3pt
At total source degrees 8 and 10, route~II also evaluates the functional without a grid. With the bodies at the foci, flat Green-function pairings of axisymmetric sources become single sums over prolate Legendre modes~\cite{Hobson:1931}. Closed expressions for the mode coefficients are obtained by rational reconstruction and recurrence inference and checked against additional modes. Summing these reconstructed expressions reproduces $\Gb_{2,6}$, $\Gb_{4,4}$ and $\Gb_{2,8}$ in exact arithmetic. $\Gb_{4,6}$ and $\Gb_{5,5}$ follow in the same way, except for eleven contributions involving incomplete elliptic integrals, which are evaluated numerically.

\vskip 3pt
\paragraph{Route III: analytic evaluation} Route~III evaluates the pairings analytically: about the field of body~1 for one to three insertions on body~2, and in total source degree for the others. In $d$ dimensions the one-body field follows from a radial recursion and Lagrange--B\"urmann inversion. The correlators $\Gb_{2,2k}$, with two insertions on the probe body, are coincident values of the Green function of the background Hessian. They are evaluated by spectral sums, true low modes plus a high-mode tail that is rational in $\ell$ and summed exactly, and the uniform tail of step~3 justifies exchanging $d\to3$ with the sum; this gives $\Gb_{2,6}$, $\Gb_{2,8}$ and $\Gb_{2,10}$. The cubic pairings give $\Gb_{3,5}$ by combining finite parts defined by spherical excisions, boundary terms and local corrections. For $\Gb_{3,7}$ they contain a rational multiple of a period $\Psi$, a double series with squared central-binomial weights, which a bivariate creative-telescoping certificate\cite{Zeilberger:1991,Chyzak:2000,Koutschan:2010} reduces to boundary terms and a single sum; Catalan's constant cancels between them, leaving $\Psi=-\frac{232}9+\frac{\pi^2}2-\pi^2\log2-\frac{35}2\zeta(3)$, whose transcendental terms cancel against those of the other contributions to $\Gb_{3,7}$. At $G^{11}$ the cubic pairings give $\Gb_{3,9}$, in agreement with the exact cubic evaluation of route~I.

\vskip 3pt
The pairings in total source degree are integrated with the two-centre Riesz formula\cite{BlanchetDamourEspositoFarese:2004} and, for the remaining terms, in three dimensions with true-sphere finite parts plus the dimensional difference, with derivatives moved onto known weights by equation~\eqref{meth:adjoint}. Hermite reduction, shuffle algebra and exact elimination with standard polylogarithm relations\cite{Brown:2009,DuhrGanglRhodes:2012} reduce the resulting functions. For $\Gb_{4,6}$ these are multiple polylogarithms of weight at most five and iterated integrals of modular forms\cite{AdamsWeinzierl:2018}, which a spectral expansion in prolate Legendre functions maps onto multiple polylogarithms at fourth roots of unity\cite{Goncharov:2001,RemiddiVermaseren:2000}. For $\Gb_{5,5}$ a M\"obius map makes them cyclotomic of level eight\cite{AblingerBlumleinSchneider:2011}, and they reduce to a rational number and a $\pi^2$ term with coefficient near $6\times10^4$, in the normalization $\Gb_{5,5}/(5!)^2$. The remaining bulk and boundary pairings cancel this term except for the $-\frac{499}{1890}\pi^2$ of the three-dimensional finite part, which the local correction removes. In both sectors every transcendental term cancels, leaving $\Gb_{4,6}=11644416$ and $\Gb_{5,5}=26428800$; the rational values emerge from the evaluation, without a rational ansatz.

\vskip 3pt
\paragraph{Momentum-space check} Without mode sums, Born series of Feynman integrals in $d=3+\epsilon$, reduced to $G$-functions with FORM\cite{Vermaseren:2000FORM}-generated numerators, give $\Gb_{2,8}$ and $\Gb_{2,10}$ from the $1/\epsilon$ residues of the summed amplitudes.

\subsection*{Results and precision of the correlators}
The 21 correlators through total source degree 12 comprise the probe tower $\Gb_{1,2k+1}=(2k)!$ of equation~\eqref{meth:probe}, the first-backreaction tower $\Gb_{2,2k}$ of equation~\eqref{eq:f2}, the established values $\Gb_{3,3}=504$, $\Gb_{3,5}=31760$ and $\Gb_{4,4}=83456$\cite{Foffa:2016rgu,DamourJaranowski:2017,Brunello:2025gpf,Brunello:2026anu}, and seven new correlators:
$\Gb_{3,7}=2814336$, $\Gb_{4,6}=11644416$, $\Gb_{5,5}=26428800$, $\Gb_{3,9}=1770222336/5$, $\Gb_{4,8}\simeq5675220992/3$, $\Gb_{5,7}\simeq6781008640$ and $\Gb_{6,6}\simeq13676344320$.

\vskip 3pt
The two towers are analytic to all orders. Route~III analytically determines the remaining correlators except $\Gb_{4,8}$, $\Gb_{5,7}$ and $\Gb_{6,6}$, whose quoted rational values are identified numerically. The exact cubic evaluation of route~I independently reproduces $\Gb_{3,7}$ and $\Gb_{3,9}$ under the assumptions stated above, and the prolate-mode sums of route~II reproduce $\Gb_{4,4}$ exactly. Analytic evaluation of the final three requires the remaining complete pairings, with the regular dimensional data retained to the order required by the local poles, including some $O(\epsilon^2)$ data for $\Gb_{4,8}$. An analytic evaluation of the higher responses would provide an alternative route.

\vskip 3pt
\paragraph{Precision} Every correlator is evaluated by at least two complementary methods, and numerical routes~I and~II cover the full set. Route~III uses no numerical field data; it shares the action, regularization prescription and source reconstruction with the numerical routes. At $G^{11}$, the numerical routes also share the exact $d$-dimensional regular data of the metric field bilinear in the sources. Comparison with the analytic $\Gb_{3,9}$ checks the use of those data in that sector. Further checks include recovery of the lower correlators without adjustment and the Green and stationarity identities.

\vskip 3pt
Route~I's per-sphere evaluations reproduce the analytic $G^9$ correlators and $\Gb_{3,9}$ to at least 19 significant digits. Its mode sums reproduce the tower $\Gb_{2,2k}$ to at least 23 significant digits, with exact agreement in the longest exact-arithmetic calculation. Route~II's grid evaluations reach 9 to 13 significant digits. Its prolate-mode evaluations, developed with the analytic values known, reach about 24 significant digits for $\Gb_{4,6}$ and $\Gb_{5,5}$. The momentum-space series reproduces $\Gb_{2,8}$ and $\Gb_{2,10}$ to about 17 significant digits.

\vskip 3pt
For the three numerically identified correlators, the per-sphere values are $\Gb_{4,8}=1891740330.66666666666671(10)$, $\Gb_{5,7}=6781008639.99999999999975(50)$ and $\Gb_{6,6}=13676344320.00000000000045(70)$, corresponding to about 22 significant digits. Parentheses give the empirical uncertainty in the last quoted digits. The relative differences from the identified rational values are $(2.3,3.7,3.3)\times10^{-23}$, within empirical relative uncertainty estimates of $(5.3,7.4,5.1)\times10^{-23}$, respectively. Route~II's grid values agree within absolute uncertainties of $(2.5$--$3)\times10^{-3}$, corresponding to about 12 significant digits. Evaluating route~I's quartic energy on route~II's grid gives $\Gb_{4,8}$ below the identified rational value by 1.7 times its own empirical uncertainty. All uncertainties are convergence estimates obtained by varying local subtractions, mode ranges and tail models, rather than rigorous interval enclosures; they are negligible on the scale of Figure~\ref{fig:mutual}.

\vskip 3pt
\paragraph{Rational identification} For the three numerically evaluated correlators, rationality is a hypothesis used in identification, not an input to the integration. Without assuming integrality, we select the rational with the smallest denominator compatible with the computed value and its empirical uncertainty. Interpreting this selection as the exact value assumes that no unrecognized systematic error exceeds the uncertainty estimate.

\vskip 3pt
Distinct rationals with denominators $D_0$ and $D$ differ by at least $1/(D_0D)$. If a computed value differs by $\delta$ from a candidate with denominator $D_0$ and has absolute empirical uncertainty $\sigma$, no other rational with $D<1/[D_0(2\sigma+\delta)]$ lies within twice that quoted uncertainty of the computed value. This criterion gives denominator thresholds of $10^{12}$, $8\times10^{11}$ and $5\times10^{11}$ for the per-sphere evaluations, compared with roughly $40$--$180$ for route~II alone. These are compatibility tests within the chosen empirical windows, not rigorous exclusions for the unknown exact values.

\vskip 3pt
The identified values share the arithmetic structure of the analytic ones: $\Gb_{4,8}/(4!\,8!)=\frac{791744}{405}$, $\Gb_{5,7}/(5!\,7!)=\frac{1513618}{135}$ and $\Gb_{6,6}/(6!)^2=\frac{3561548}{135}$ have denominators dividing $14175=3^4\cdot5^2\cdot7$, the denominator of the analytic $\Gb_{3,9}/(3!\,9!)=\frac{2304977}{14175}$. The comparisons were not blind: the per-sphere evaluations at $G^{11}$ were developed with earlier values known, and the reported uncertainty of $\Gb_{5,7}$ was set after comparison.

\vskip 3pt
The exact source reconstruction propagates the same three numerical inputs into both mass representations, so their effects are correlated. These inputs supply 4.8--6.5\% of the affected coefficients of equation~\eqref{eq:fullpotential} at $G^{11}$ and 9.6--12.5\% at $G^{12}$. The coefficients determined analytically carry no numerical uncertainty.

\subsection*{The potential in symmetric mass variables}
The symmetric variables used in Figure~\ref{fig:mutual} reorganize the same potential; the change of variables alters both its normalization and the mass held fixed in the expansion in the mass ratio.

\vskip 3pt
Equation~\eqref{eq:fullpotential} and Figure~\ref{fig:coverage} use $V/m_2=\sum_n\alpha^n\mathscr Q_n(q)$, with $\mathscr Q_n(q)=\sum_{k=0}^{n-1}c_{n-k,k+1}q^k$ in terms of the coefficients of equation~\eqref{meth:coefficient}. Exchange symmetry gives $q^{n-1}\mathscr Q_n(1/q)=\mathscr Q_n(q)$. Figure~\ref{fig:mutual} uses $M=m_1+m_2$, $\mu=m_1m_2/M$, $\nu=\mu/M=q/(1+q)^2$ and $u=GM/r=(1+q)\alpha$, which are unchanged by $m_1\leftrightarrow m_2$, with $0\le\nu\le\frac14$. The coefficient of $G^n/r^n$ in $V$, of mass degree $n+1$, is $m_1m_2$ times a symmetric polynomial of degree $n-1$, hence a polynomial in $M$ and $m_1m_2=\nu M^2$; thus $V/\mu=\sum_nu^n\mathscr P_n(\nu)$, with $\mathscr P_n=\sum_kp_{n,k}\nu^k$ of degree at most $\lfloor(n-1)/2\rfloor$, $\mathscr Q_n(q)=(1+q)^{n-1}\mathscr P_n(q/(1+q)^2)$ and
\begin{equation}
 c_{a,n+1-a}=\sum_{k=0}^{\lfloor(n-1)/2\rfloor}p_{n,k}\binom{n-1-2k}{a-k-1},
 \label{meth:massbasis}
\end{equation}
a relation triangular with unit diagonal for $a=1,\ldots,\lfloor(n+1)/2\rfloor$. Hence $p_{n,k}$ depends only on correlators with $\min(a,b)\le k+1$: the coefficients of $\nu^0$ to $\nu^2$ are exact through $G^{12}$, and those of $\nu^3$ to $\nu^5$ at $G^{11}$ and $G^{12}$ rest on the identified $\Gb_{4,8}$, $\Gb_{5,7}$ and $\Gb_{6,6}$. The degree bound mirrors the mass polynomiality of the post-Minkowskian scattering angle\cite{Damour:2019lcq}, which, combined with self-force results, constrains post-Newtonian dynamics\cite{Bini:2019nra}. The complete polynomials are
\begin{align}
 &\mathscr P_1=-1, ~\qquad\qquad\qquad \mathscr P_2=\tfrac12,
 \nonumber\\[0.75 em]
  &\mathscr P_3=-\tfrac12-2\nu, \qquad\qquad \mathscr P_4=\tfrac38+\tfrac{39}{8}\nu,
 \nonumber\\[0.75 em]
 &\mathscr P_5=-\tfrac38-\tfrac{53}{6}\nu-\tfrac{46}{3}\nu^2,\qquad
 \nonumber\\[0.75 em]
 &\mathscr P_6=\tfrac5{16}+\tfrac{653}{48}\nu+\tfrac{3115}{48}\nu^2,
 \nonumber\\[0.75 em]
 &\mathscr P_7=-\tfrac5{16}-\tfrac{1385}{72}\nu-\tfrac{1079}{6}\nu^2-\tfrac{1520}{9}\nu^3,\qquad
 \nonumber\\[0.75 em]
 &\mathscr P_8=\tfrac{35}{128}+\tfrac{29539}{1152}\nu+\tfrac{227837}{576}\nu^2+\tfrac{1186819}{1152}\nu^3,
 \nonumber\\[0.75 em]
 &\mathscr P_9=-\tfrac{35}{128}-\tfrac{23609}{720}\nu-\tfrac{136651}{180}\nu^2-\tfrac{344221}{90}\nu^3-\tfrac{6746}{3}\nu^4,\qquad
 \\[0.75 em]
 &
 \mathscr P_{10}=\tfrac{63}{256}+\tfrac{52077}{1280}\nu+\tfrac{338485}{256}\nu^2+\tfrac{62808347}{5760}\nu^3+\tfrac{68909689}{3840}\nu^4,
 \nonumber\\[0.75 em]
 &\mathscr P_{11}=-\tfrac{63}{256}-\tfrac{661837}{13440}\nu-\tfrac{487120337}{226800}\nu^2-\tfrac{1491989017}{56700}\nu^3-\tfrac{2369253391}{28350}\nu^4-\tfrac{476299456}{14175}\nu^5,
  \nonumber\\[0.75 em]
 &\mathscr P_{12}=\tfrac{231}{1024}+\tfrac{6292939}{107520}\nu+\tfrac{443964293}{134400}\nu^2+\tfrac{818853576619}{14515200}\nu^3+\tfrac{1413221816597}{4838400}\nu^4+\tfrac{4813187181437}{14515200}\nu^5 .
  \nonumber
\label{meth:symmetricpolys}
\end{align}
Through the computed order $G^{12}$, all nonzero $c_{a,b}$ at $a+b=n+1$ and $p_{n,k}$ have sign $(-1)^n$, so $|\mathscr P_n(\nu)|=\sum_k|p_{n,k}|\nu^k$ for $\nu\ge0$. The factor $(1+q)^{n-1}$ reflects the combined change of coupling and energy normalization. At equal masses, $\mathscr Q_n(1)=2^{n-1}\mathscr P_n(\frac14)$. Figure~\ref{fig:mutual} uses $u$ to compare coefficients at fixed total mass. Substituting $\alpha=u/(1+q)$ and $\mu=m_2/(1+q)$ into $V/m_2=\Phi_0(\alpha)+q\Phi_1(\alpha)+O(q^2)$ and using $\nu=q+O(q^2)$ gives $V/\mu=\Phi_0(u)+\nu\Phi_1^{\rm sym}(u)+O(\nu^2)$, with
\begin{equation}
 \Phi_1^{\rm sym}(u)=\Phi_0(u)-u\Phi_0'(u)+\Phi_1(u)=\frac{1+u-u^2}{(1+u)\sqrt{1-u^2}}-1+\Phi_1(u)=-2u^3+\tfrac{39}8u^4-\tfrac{53}6u^5+O(u^6);
 \label{meth:symmetricSF}
\end{equation}
the changes of normalization and coupling cancel the $u^2/2$ term of $\Phi_1$. Through $G^{12}$, $\mathscr P_n(0)$ and $\mathscr P_n'(0)$ (the prime denoting $\dd/\dd\nu$) are the Taylor coefficients of $\Phi_0$ and $\Phi_1^{\rm sym}$. The two mass representations agree at every finite order, but their closed 0SF and 1SF sectors sum different terms beyond $G^{12}$.

\vskip 3pt
\paragraph{Growth of the coefficients} At equal masses the ratio $|\mathscr P_n(\frac14)/\mathscr P_{n-1}(\frac14)|$ rises from 2.19 at $n=6$ to 2.80 at $n=12$, where $\mathscr P_{12}(\frac14)\simeq2.57\times10^3$; the $\nu^0$ and $\nu^1$ terms supply 0.58\% of this coefficient, and the $\nu^3$ and $\nu^4$ terms 79\%. At $G^{12}$ the terms of second and higher order in $\nu$ exceed the 0SF and 1SF terms for $\nu\gtrsim0.017$. Ratio analysis\cite{DombSykes:1957,Guttmann:1989} and Pad\'e approximants\cite{BakerGravesMorris:1996} of the known orders place the nearest singularity of the series at fixed $\nu$ on the negative real axis for every $\nu$ examined, between about $u=-0.36$ and $-0.28$ for equal masses. These estimates are heuristic, depend on the coordinate convention and do not establish the asymptotic behaviour of the uncomputed coefficients.

\vskip 3pt
\paragraph{Share of the first-order term} The closed functions and finite orders combine without double counting. Define $H_{12}(u,\nu)=\sum_{n=1}^{12}u^n[\mathscr P_n(\nu)-\mathscr P_n(0)-\nu\mathscr P_n'(0)]$. Restoring $c$ in the energy normalization, the potential $\mu c^2[\Phi_0(u)+\nu\Phi_1^{\rm sym}(u)+H_{12}]$ retains every term through $G^{12}$ and the 0SF and 1SF terms to all orders, with $u=GM/(rc^2)$. The omitted series begins at $u^{13}$ and is at least quadratic in $\nu$.

\vskip 3pt
Figure~\ref{fig:mutual} (right) shows $s_1=\nu\Phi_1^{\rm sym}/(\nu\Phi_1^{\rm sym}+H_{12})$. At fixed $\nu$, its weak-field expansion is $s_1=1-\frac{23}3\nu u^2+O(u^3)$. The bands vary the denominator by $\pm u^{12}[\mathscr P_{12}(\nu)-\mathscr P_{12}(0)-\nu\mathscr P_{12}'(0)]$, the last retained higher-mass-ratio contribution. This is a truncation diagnostic, not an error bound. At $u=0.2$ the shares, with the band half-widths in parentheses, are $0.9887(1)$, $0.9774(5)$ and $0.944(4)$ for $\nu=0.05$, $0.1$ and $\frac14$. For equal masses, the magnitude of the $G^{12}$ term is about $0.56$ times that of the $G^{11}$ term.

\vskip 3pt
These fractions depend on the gauge convention and mass variables. At $\lambda=0$, $\Phi_1^{\rm sym}(0.2)=+0.0011$ instead of $-0.0104$; in the fixed-$m_1$ variables of equation~\eqref{eq:fullpotential}, at the same separation and $\lambda=1$, $q\Phi_1$ carries $107\%$ of the equal-mass correction beyond the probe term. Thus $s_1$ characterizes the potential in the stated convention and total-mass variables. The value $u=0.2$ is a reference coupling, since the static potential alone does not define an orbit. Self-force studies of comparable masses compare gauge-invariant orbital quantities instead~\cite{LeTiecEtAl:2011,PoundEtAl:2020,vandeMeent:2020xgc,WardellEtAl:2023}.

\subsection*{The complete potential in connected correlators}
Equations~\eqref{meth:GammaG1}--\eqref{meth:GammaG12} give the reconstruction of equation~\eqref{meth:gammaexpansion} explicitly through $G^{12}$, extending the organization of Appendix~A of~\cite{Brunello:2026anu}, which gives it through $G^8$. We suppress the argument $r$ of the dimensional correlators $\Gamma_{a,b}(r)$ of equation~\eqref{meth:gammadefinition}. Unequal mass powers are paired with their exchanged partners, and equal powers occur once. At odd orders each bracket ends with the correlator of its own mass sector, with unit coefficient; all other terms are products of lower correlators generated by the particle couplings. At even orders no new correlator enters. Substituting the 21 correlators computed in this work gives equation~\eqref{eq:fullpotential}.
\begingroup
\allowdisplaybreaks[4]
\setlength{\jot}{1pt}
\begin{align}
V_{G^{1}}={}&-m_1m_2\Gamma_{1,1},\label{meth:GammaG1}\\[3pt]
V_{G^{2}}={}&\frac{m_1m_2^2+m_1^2m_2}{2}\Gamma_{1,1}^2,\label{meth:GammaG2}\\[3pt]
V_{G^{3}}={}&-\Biggl\{\frac{m_1m_2^{3}+m_1^{3}m_2}{6}\Bigl(\Gamma_{1,1}^{3}+\Gamma_{1,3}\Bigr)+\frac{m_1^{2}m_2^{2}}{4}\Bigl(4\Gamma_{1,1}^{3}+\Gamma_{2,2}\Bigr)\Biggr\},\label{meth:GammaG3}\\[3pt]
V_{G^{4}}={}&\Biggl\{\frac{m_1m_2^{4}+m_1^{4}m_2}{24}\Bigl(\Gamma_{1,1}^{4}+4\Gamma_{1,1}\Gamma_{1,3}\Bigr)+\frac{m_1^{2}m_2^{3}+m_1^{3}m_2^{2}}{2}\Bigl(2\Gamma_{1,1}^{4}+\Gamma_{1,1}\Gamma_{1,3}+\Gamma_{1,1}\Gamma_{2,2}\Bigr)\Biggr\},\label{meth:GammaG4}\\[3pt]
V_{G^{5}}={}&-\Biggl\{\frac{m_1m_2^{5}+m_1^{5}m_2}{120}\Bigl(\Gamma_{1,1}^{5}+10\Gamma_{1,1}^{2}\Gamma_{1,3}+\Gamma_{1,5}\Bigr)\label{meth:GammaG5}\\*
&\quad+\frac{m_1^{2}m_2^{4}+m_1^{4}m_2^{2}}{48}\Bigl(32\Gamma_{1,1}^{5}+56\Gamma_{1,1}^{2}\Gamma_{1,3}+24\Gamma_{1,1}^{2}\Gamma_{2,2}+\Gamma_{2,4}\Bigr)\notag\\
&\quad+\frac{m_1^{3}m_2^{3}}{36}\Bigl(81\Gamma_{1,1}^{5}+54\Gamma_{1,1}^{2}\Gamma_{1,3}+72\Gamma_{1,1}^{2}\Gamma_{2,2}+\Gamma_{3,3}\Bigr)\Biggr\},\notag\\[3pt]
V_{G^{6}}={}&\Biggl\{\frac{m_1m_2^{6}+m_1^{6}m_2}{720}\Bigl(\Gamma_{1,1}^{6}+20\Gamma_{1,1}^{3}\Gamma_{1,3}+6\Gamma_{1,1}\Gamma_{1,5}+10\Gamma_{1,3}^{2}\Bigr)\label{meth:GammaG6}\\*
&\quad+\frac{m_1^{2}m_2^{5}+m_1^{5}m_2^{2}}{24}\Bigl(8\Gamma_{1,1}^{6}+32\Gamma_{1,1}^{3}\Gamma_{1,3}+8\Gamma_{1,1}^{3}\Gamma_{2,2}+\Gamma_{1,1}\Gamma_{1,5}+\Gamma_{1,1}\Gamma_{2,4}+3\Gamma_{1,3}^{2}+2\Gamma_{1,3}\Gamma_{2,2}\Bigr)\notag\\
&\quad+\frac{m_1^{3}m_2^{4}+m_1^{4}m_2^{3}}{24}\Bigl(72\Gamma_{1,1}^{6}+98\Gamma_{1,1}^{3}\Gamma_{1,3}+90\Gamma_{1,1}^{3}\Gamma_{2,2}+2\Gamma_{1,1}\Gamma_{2,4}+2\Gamma_{1,1}\Gamma_{3,3}+2\Gamma_{1,3}^{2}+6\Gamma_{1,3}\Gamma_{2,2}+3\Gamma_{2,2}^{2}\Bigr)\Biggr\},\notag\\[3pt]
V_{G^{7}}={}&-\Biggl\{\frac{m_1m_2^{7}+m_1^{7}m_2}{5040}\Bigl(\Gamma_{1,1}^{7}+35\Gamma_{1,1}^{4}\Gamma_{1,3}+21\Gamma_{1,1}^{2}\Gamma_{1,5}+70\Gamma_{1,1}\Gamma_{1,3}^{2}+\Gamma_{1,7}\Bigr)\label{meth:GammaG7}\\*
&\quad+\frac{m_1^{2}m_2^{6}+m_1^{6}m_2^{2}}{1440}\Bigl(192\Gamma_{1,1}^{7}+1440\Gamma_{1,1}^{4}\Gamma_{1,3}+240\Gamma_{1,1}^{4}\Gamma_{2,2}+132\Gamma_{1,1}^{2}\Gamma_{1,5}+60\Gamma_{1,1}^{2}\Gamma_{2,4}\notag\\
&\qquad{}+600\Gamma_{1,1}\Gamma_{1,3}^{2}+240\Gamma_{1,1}\Gamma_{1,3}\Gamma_{2,2}+\Gamma_{2,6}\Bigr)\notag\\
&\quad+\frac{m_1^{3}m_2^{5}+m_1^{5}m_2^{3}}{720}\Bigl(2025\Gamma_{1,1}^{7}+5355\Gamma_{1,1}^{4}\Gamma_{1,3}+3240\Gamma_{1,1}^{4}\Gamma_{2,2}+75\Gamma_{1,1}^{2}\Gamma_{1,5}+210\Gamma_{1,1}^{2}\Gamma_{2,4}\notag\\
&\qquad{}+90\Gamma_{1,1}^{2}\Gamma_{3,3}+630\Gamma_{1,1}\Gamma_{1,3}^{2}+1080\Gamma_{1,1}\Gamma_{1,3}\Gamma_{2,2}+270\Gamma_{1,1}\Gamma_{2,2}^{2}+\Gamma_{3,5}\Bigr)\notag\\
&\quad+\frac{m_1^{4}m_2^{4}}{576}\Bigl(4096\Gamma_{1,1}^{7}+6656\Gamma_{1,1}^{4}\Gamma_{1,3}+6912\Gamma_{1,1}^{4}\Gamma_{2,2}+192\Gamma_{1,1}^{2}\Gamma_{2,4}+240\Gamma_{1,1}^{2}\Gamma_{3,3}\notag\\
&\qquad{}+544\Gamma_{1,1}\Gamma_{1,3}^{2}+1152\Gamma_{1,1}\Gamma_{1,3}\Gamma_{2,2}+720\Gamma_{1,1}\Gamma_{2,2}^{2}+\Gamma_{4,4}\Bigr)\Biggr\},\notag\\[3pt]
V_{G^{8}}={}&\Biggl\{\frac{m_1m_2^{8}+m_1^{8}m_2}{40320}\Bigl(\Gamma_{1,1}^{8}+56\Gamma_{1,1}^{5}\Gamma_{1,3}+56\Gamma_{1,1}^{3}\Gamma_{1,5}+280\Gamma_{1,1}^{2}\Gamma_{1,3}^{2}+8\Gamma_{1,1}\Gamma_{1,7}+56\Gamma_{1,3}\Gamma_{1,5}\Bigr)\label{meth:GammaG8}\\*
&\quad+\frac{m_1^{2}m_2^{7}+m_1^{7}m_2^{2}}{720}\Bigl(32\Gamma_{1,1}^{8}+400\Gamma_{1,1}^{5}\Gamma_{1,3}+48\Gamma_{1,1}^{5}\Gamma_{2,2}+72\Gamma_{1,1}^{3}\Gamma_{1,5}+20\Gamma_{1,1}^{3}\Gamma_{2,4}+440\Gamma_{1,1}^{2}\Gamma_{1,3}^{2}\notag\\
&\qquad{}+120\Gamma_{1,1}^{2}\Gamma_{1,3}\Gamma_{2,2}+\Gamma_{1,1}\Gamma_{1,7}+\Gamma_{1,1}\Gamma_{2,6}+15\Gamma_{1,3}\Gamma_{1,5}+5\Gamma_{1,3}\Gamma_{2,4}+3\Gamma_{2,2}\Gamma_{1,5}\Bigr)\notag\\
&\quad+\frac{m_1^{3}m_2^{6}+m_1^{6}m_2^{3}}{720}\Bigl(1458\Gamma_{1,1}^{8}+6723\Gamma_{1,1}^{5}\Gamma_{1,3}+2835\Gamma_{1,1}^{5}\Gamma_{2,2}+288\Gamma_{1,1}^{3}\Gamma_{1,5}+360\Gamma_{1,1}^{3}\Gamma_{2,4}\notag\\
&\qquad{}+90\Gamma_{1,1}^{3}\Gamma_{3,3}+2430\Gamma_{1,1}^{2}\Gamma_{1,3}^{2}+2610\Gamma_{1,1}^{2}\Gamma_{1,3}\Gamma_{2,2}+405\Gamma_{1,1}^{2}\Gamma_{2,2}^{2}+3\Gamma_{1,1}\Gamma_{2,6}+3\Gamma_{1,1}\Gamma_{3,5}\notag\\
&\qquad{}+3\Gamma_{1,3}\Gamma_{1,5}+30\Gamma_{1,3}\Gamma_{2,4}+10\Gamma_{1,3}\Gamma_{3,3}+15\Gamma_{2,2}\Gamma_{1,5}+15\Gamma_{2,2}\Gamma_{2,4}\Bigr)\notag\\
&\quad+\frac{m_1^{4}m_2^{5}+m_1^{5}m_2^{4}}{144}\Bigl(1600\Gamma_{1,1}^{8}+3824\Gamma_{1,1}^{5}\Gamma_{1,3}+3360\Gamma_{1,1}^{5}\Gamma_{2,2}+25\Gamma_{1,1}^{3}\Gamma_{1,5}+167\Gamma_{1,1}^{3}\Gamma_{2,4}\notag\\
&\qquad{}+144\Gamma_{1,1}^{3}\Gamma_{3,3}+721\Gamma_{1,1}^{2}\Gamma_{1,3}^{2}+1398\Gamma_{1,1}^{2}\Gamma_{1,3}\Gamma_{2,2}+648\Gamma_{1,1}^{2}\Gamma_{2,2}^{2}+\Gamma_{1,1}\Gamma_{3,5}+\Gamma_{1,1}\Gamma_{4,4}\notag\\
&\qquad{}+\Gamma_{1,3}\Gamma_{1,5}+5\Gamma_{1,3}\Gamma_{2,4}+6\Gamma_{1,3}\Gamma_{3,3}+6\Gamma_{2,2}\Gamma_{2,4}+6\Gamma_{2,2}\Gamma_{3,3}\Bigr)\Biggr\},\notag\\[3pt]
V_{G^{9}}={}&-\Biggl\{\frac{m_1m_2^{9}+m_1^{9}m_2}{362880}\Bigl(\Gamma_{1,1}^{9}+84\Gamma_{1,1}^{6}\Gamma_{1,3}+126\Gamma_{1,1}^{4}\Gamma_{1,5}+840\Gamma_{1,1}^{3}\Gamma_{1,3}^{2}+36\Gamma_{1,1}^{2}\Gamma_{1,7}\label{meth:GammaG9}\\*
&\qquad{}+504\Gamma_{1,1}\Gamma_{1,3}\Gamma_{1,5}+280\Gamma_{1,3}^{3}+\Gamma_{1,9}\Bigr)\notag\\
&\quad+\frac{m_1^{2}m_2^{8}+m_1^{8}m_2^{2}}{80640}\Bigl(1024\Gamma_{1,1}^{9}+19712\Gamma_{1,1}^{6}\Gamma_{1,3}+1792\Gamma_{1,1}^{6}\Gamma_{2,2}+5824\Gamma_{1,1}^{4}\Gamma_{1,5}+1120\Gamma_{1,1}^{4}\Gamma_{2,4}\notag\\
&\qquad{}+44800\Gamma_{1,1}^{3}\Gamma_{1,3}^{2}+8960\Gamma_{1,1}^{3}\Gamma_{1,3}\Gamma_{2,2}+240\Gamma_{1,1}^{2}\Gamma_{1,7}+112\Gamma_{1,1}^{2}\Gamma_{2,6}+5152\Gamma_{1,1}\Gamma_{1,3}\Gamma_{1,5}\notag\\
&\qquad{}+1120\Gamma_{1,1}\Gamma_{1,3}\Gamma_{2,4}+672\Gamma_{1,1}\Gamma_{2,2}\Gamma_{1,5}+3360\Gamma_{1,3}^{3}+1120\Gamma_{1,3}^{2}\Gamma_{2,2}+\Gamma_{2,8}\Bigr)\notag\\
&\quad+\frac{m_1^{3}m_2^{7}+m_1^{7}m_2^{3}}{30240}\Bigl(35721\Gamma_{1,1}^{9}+263088\Gamma_{1,1}^{6}\Gamma_{1,3}+81648\Gamma_{1,1}^{6}\Gamma_{2,2}+22491\Gamma_{1,1}^{4}\Gamma_{1,5}+17010\Gamma_{1,1}^{4}\Gamma_{2,4}\notag\\
&\qquad{}+2835\Gamma_{1,1}^{4}\Gamma_{3,3}+214200\Gamma_{1,1}^{3}\Gamma_{1,3}^{2}+158760\Gamma_{1,1}^{3}\Gamma_{1,3}\Gamma_{2,2}+17010\Gamma_{1,1}^{3}\Gamma_{2,2}^{2}+147\Gamma_{1,1}^{2}\Gamma_{1,7}+420\Gamma_{1,1}^{2}\Gamma_{2,6}\notag\\
&\qquad{}+189\Gamma_{1,1}^{2}\Gamma_{3,5}+4788\Gamma_{1,1}\Gamma_{1,3}\Gamma_{1,5}+6300\Gamma_{1,1}\Gamma_{1,3}\Gamma_{2,4}+1260\Gamma_{1,1}\Gamma_{1,3}\Gamma_{3,3}+3528\Gamma_{1,1}\Gamma_{2,2}\Gamma_{1,5}\notag\\
&\qquad{}+1890\Gamma_{1,1}\Gamma_{2,2}\Gamma_{2,4}+5040\Gamma_{1,3}^{3}+7560\Gamma_{1,3}^{2}\Gamma_{2,2}+1890\Gamma_{1,3}\Gamma_{2,2}^{2}+\Gamma_{3,7}\Bigr)\notag\\
&\quad+\frac{m_1^{4}m_2^{6}+m_1^{6}m_2^{4}}{17280}\Bigl(221184\Gamma_{1,1}^{9}+807936\Gamma_{1,1}^{6}\Gamma_{1,3}+552960\Gamma_{1,1}^{6}\Gamma_{2,2}+18144\Gamma_{1,1}^{4}\Gamma_{1,5}+47040\Gamma_{1,1}^{4}\Gamma_{2,4}\notag\\
&\qquad{}+26880\Gamma_{1,1}^{4}\Gamma_{3,3}+312960\Gamma_{1,1}^{3}\Gamma_{1,3}^{2}+495360\Gamma_{1,1}^{3}\Gamma_{1,3}\Gamma_{2,2}+161280\Gamma_{1,1}^{3}\Gamma_{2,2}^{2}+216\Gamma_{1,1}^{2}\Gamma_{2,6}+552\Gamma_{1,1}^{2}\Gamma_{3,5}\notag\\
&\qquad{}+240\Gamma_{1,1}^{2}\Gamma_{4,4}+1296\Gamma_{1,1}\Gamma_{1,3}\Gamma_{1,5}+6720\Gamma_{1,1}\Gamma_{1,3}\Gamma_{2,4}+5280\Gamma_{1,1}\Gamma_{1,3}\Gamma_{3,3}+2160\Gamma_{1,1}\Gamma_{2,2}\Gamma_{1,5}\notag\\
&\qquad{}+6120\Gamma_{1,1}\Gamma_{2,2}\Gamma_{2,4}+2880\Gamma_{1,1}\Gamma_{2,2}\Gamma_{3,3}+3120\Gamma_{1,3}^{3}+9360\Gamma_{1,3}^{2}\Gamma_{2,2}+7920\Gamma_{1,3}\Gamma_{2,2}^{2}+1440\Gamma_{2,2}^{3}+\Gamma_{4,6}\Bigr)\notag\\
&\quad+\frac{m_1^{5}m_2^{5}}{14400}\Bigl(390625\Gamma_{1,1}^{9}+1062500\Gamma_{1,1}^{6}\Gamma_{1,3}+1000000\Gamma_{1,1}^{6}\Gamma_{2,2}+6250\Gamma_{1,1}^{4}\Gamma_{1,5}+55000\Gamma_{1,1}^{4}\Gamma_{2,4}\notag\\
&\qquad{}+52500\Gamma_{1,1}^{4}\Gamma_{3,3}+325000\Gamma_{1,1}^{3}\Gamma_{1,3}^{2}+620000\Gamma_{1,1}^{3}\Gamma_{1,3}\Gamma_{2,2}+315000\Gamma_{1,1}^{3}\Gamma_{2,2}^{2}+500\Gamma_{1,1}^{2}\Gamma_{3,5}\notag\\
&\qquad{}+600\Gamma_{1,1}^{2}\Gamma_{4,4}+1000\Gamma_{1,1}\Gamma_{1,3}\Gamma_{1,5}+6400\Gamma_{1,1}\Gamma_{1,3}\Gamma_{2,4}+6000\Gamma_{1,1}\Gamma_{1,3}\Gamma_{3,3}+6000\Gamma_{1,1}\Gamma_{2,2}\Gamma_{2,4}\notag\\
&\qquad{}+7200\Gamma_{1,1}\Gamma_{2,2}\Gamma_{3,3}+3000\Gamma_{1,3}^{3}+8800\Gamma_{1,3}^{2}\Gamma_{2,2}+9000\Gamma_{1,3}\Gamma_{2,2}^{2}+3600\Gamma_{2,2}^{3}+\Gamma_{5,5}\Bigr)\Biggr\},\notag\\[3pt]
V_{G^{10}}={}&\Biggl\{\frac{m_1m_2^{10}+m_1^{10}m_2}{3628800}\Bigl(\Gamma_{1,1}^{10}+120\Gamma_{1,1}^{7}\Gamma_{1,3}+252\Gamma_{1,1}^{5}\Gamma_{1,5}+2100\Gamma_{1,1}^{4}\Gamma_{1,3}^{2}+120\Gamma_{1,1}^{3}\Gamma_{1,7}\label{meth:GammaG10}\\*
&\qquad{}+2520\Gamma_{1,1}^{2}\Gamma_{1,3}\Gamma_{1,5}+2800\Gamma_{1,1}\Gamma_{1,3}^{3}+10\Gamma_{1,1}\Gamma_{1,9}+120\Gamma_{1,3}\Gamma_{1,7}+126\Gamma_{1,5}^{2}\Bigr)\notag\\
&\quad+\frac{m_1^{2}m_2^{9}+m_1^{9}m_2^{2}}{120960}\Bigl(384\Gamma_{1,1}^{10}+10752\Gamma_{1,1}^{7}\Gamma_{1,3}+768\Gamma_{1,1}^{7}\Gamma_{2,2}+4704\Gamma_{1,1}^{5}\Gamma_{1,5}+672\Gamma_{1,1}^{5}\Gamma_{2,4}\notag\\
&\qquad{}+43680\Gamma_{1,1}^{4}\Gamma_{1,3}^{2}+6720\Gamma_{1,1}^{4}\Gamma_{1,3}\Gamma_{2,2}+384\Gamma_{1,1}^{3}\Gamma_{1,7}+112\Gamma_{1,1}^{3}\Gamma_{2,6}+10752\Gamma_{1,1}^{2}\Gamma_{1,3}\Gamma_{1,5}\notag\\
&\qquad{}+1680\Gamma_{1,1}^{2}\Gamma_{1,3}\Gamma_{2,4}+1008\Gamma_{1,1}^{2}\Gamma_{2,2}\Gamma_{1,5}+13440\Gamma_{1,1}\Gamma_{1,3}^{3}+3360\Gamma_{1,1}\Gamma_{1,3}^{2}\Gamma_{2,2}+3\Gamma_{1,1}\Gamma_{1,9}\notag\\
&\qquad{}+3\Gamma_{1,1}\Gamma_{2,8}+84\Gamma_{1,3}\Gamma_{1,7}+28\Gamma_{1,3}\Gamma_{2,6}+12\Gamma_{2,2}\Gamma_{1,7}+105\Gamma_{1,5}^{2}+42\Gamma_{1,5}\Gamma_{2,4}\Bigr)\notag\\
&\quad+\frac{m_1^{3}m_2^{8}+m_1^{8}m_2^{3}}{40320}\Bigl(23328\Gamma_{1,1}^{10}+256932\Gamma_{1,1}^{7}\Gamma_{1,3}+61236\Gamma_{1,1}^{7}\Gamma_{2,2}+36288\Gamma_{1,1}^{5}\Gamma_{1,5}+18900\Gamma_{1,1}^{5}\Gamma_{2,4}\notag\\
&\qquad{}+2268\Gamma_{1,1}^{5}\Gamma_{3,3}+394380\Gamma_{1,1}^{4}\Gamma_{1,3}^{2}+215460\Gamma_{1,1}^{4}\Gamma_{1,3}\Gamma_{2,2}+17010\Gamma_{1,1}^{4}\Gamma_{2,2}^{2}+704\Gamma_{1,1}^{3}\Gamma_{1,7}+924\Gamma_{1,1}^{3}\Gamma_{2,6}\notag\\
&\qquad{}+252\Gamma_{1,1}^{3}\Gamma_{3,5}+26292\Gamma_{1,1}^{2}\Gamma_{1,3}\Gamma_{1,5}+18480\Gamma_{1,1}^{2}\Gamma_{1,3}\Gamma_{2,4}+2520\Gamma_{1,1}^{2}\Gamma_{1,3}\Gamma_{3,3}+10836\Gamma_{1,1}^{2}\Gamma_{2,2}\Gamma_{1,5}\notag\\
&\qquad{}+3780\Gamma_{1,1}^{2}\Gamma_{2,2}\Gamma_{2,4}+42840\Gamma_{1,1}\Gamma_{1,3}^{3}+42840\Gamma_{1,1}\Gamma_{1,3}^{2}\Gamma_{2,2}+7560\Gamma_{1,1}\Gamma_{1,3}\Gamma_{2,2}^{2}+4\Gamma_{1,1}\Gamma_{2,8}+4\Gamma_{1,1}\Gamma_{3,7}\notag\\
&\qquad{}+4\Gamma_{1,3}\Gamma_{1,7}+84\Gamma_{1,3}\Gamma_{2,6}+28\Gamma_{1,3}\Gamma_{3,5}+28\Gamma_{2,2}\Gamma_{1,7}+28\Gamma_{2,2}\Gamma_{2,6}+140\Gamma_{1,5}\Gamma_{2,4}+28\Gamma_{1,5}\Gamma_{3,3}+35\Gamma_{2,4}^{2}\Bigr)\notag\\
&\quad+\frac{m_1^{4}m_2^{7}+m_1^{7}m_2^{4}}{4320}\Bigl(50176\Gamma_{1,1}^{10}+273408\Gamma_{1,1}^{7}\Gamma_{1,3}+145152\Gamma_{1,1}^{7}\Gamma_{2,2}+12936\Gamma_{1,1}^{5}\Gamma_{1,5}+19248\Gamma_{1,1}^{5}\Gamma_{2,4}\notag\\
&\qquad{}+7680\Gamma_{1,1}^{5}\Gamma_{3,3}+198360\Gamma_{1,1}^{4}\Gamma_{1,3}^{2}+243360\Gamma_{1,1}^{4}\Gamma_{1,3}\Gamma_{2,2}+57600\Gamma_{1,1}^{4}\Gamma_{2,2}^{2}+49\Gamma_{1,1}^{3}\Gamma_{1,7}+273\Gamma_{1,1}^{3}\Gamma_{2,6}\notag\\
&\qquad{}+312\Gamma_{1,1}^{3}\Gamma_{3,5}+80\Gamma_{1,1}^{3}\Gamma_{4,4}+3333\Gamma_{1,1}^{2}\Gamma_{1,3}\Gamma_{1,5}+8025\Gamma_{1,1}^{2}\Gamma_{1,3}\Gamma_{2,4}+4080\Gamma_{1,1}^{2}\Gamma_{1,3}\Gamma_{3,3}\notag\\
&\qquad{}+3717\Gamma_{1,1}^{2}\Gamma_{2,2}\Gamma_{1,5}+5040\Gamma_{1,1}^{2}\Gamma_{2,2}\Gamma_{2,4}+1440\Gamma_{1,1}^{2}\Gamma_{2,2}\Gamma_{3,3}+9130\Gamma_{1,1}\Gamma_{1,3}^{3}+21150\Gamma_{1,1}\Gamma_{1,3}^{2}\Gamma_{2,2}\notag\\
&\qquad{}+12240\Gamma_{1,1}\Gamma_{1,3}\Gamma_{2,2}^{2}+1440\Gamma_{1,1}\Gamma_{2,2}^{3}+\Gamma_{1,1}\Gamma_{3,7}+\Gamma_{1,1}\Gamma_{4,6}+\Gamma_{1,3}\Gamma_{1,7}+3\Gamma_{1,3}\Gamma_{2,6}+15\Gamma_{1,3}\Gamma_{3,5}\notag\\
&\qquad{}+5\Gamma_{1,3}\Gamma_{4,4}+9\Gamma_{2,2}\Gamma_{2,6}+9\Gamma_{2,2}\Gamma_{3,5}+3\Gamma_{1,5}\Gamma_{2,4}+15\Gamma_{1,5}\Gamma_{3,3}+15\Gamma_{2,4}^{2}+15\Gamma_{2,4}\Gamma_{3,3}\Bigr)\notag\\
&\quad+\frac{m_1^{5}m_2^{6}+m_1^{6}m_2^{5}}{8640}\Bigl(405000\Gamma_{1,1}^{10}+1422000\Gamma_{1,1}^{7}\Gamma_{1,3}+1215000\Gamma_{1,1}^{7}\Gamma_{2,2}+18723\Gamma_{1,1}^{5}\Gamma_{1,5}+89175\Gamma_{1,1}^{5}\Gamma_{2,4}\notag\\
&\qquad{}+72000\Gamma_{1,1}^{5}\Gamma_{3,3}+686925\Gamma_{1,1}^{4}\Gamma_{1,3}^{2}+1223550\Gamma_{1,1}^{4}\Gamma_{1,3}\Gamma_{2,2}+540000\Gamma_{1,1}^{4}\Gamma_{2,2}^{2}+216\Gamma_{1,1}^{3}\Gamma_{2,6}\notag\\
&\qquad{}+1258\Gamma_{1,1}^{3}\Gamma_{3,5}+1050\Gamma_{1,1}^{3}\Gamma_{4,4}+4350\Gamma_{1,1}^{2}\Gamma_{1,3}\Gamma_{1,5}+23430\Gamma_{1,1}^{2}\Gamma_{1,3}\Gamma_{2,4}+20340\Gamma_{1,1}^{2}\Gamma_{1,3}\Gamma_{3,3}\notag\\
&\qquad{}+3240\Gamma_{1,1}^{2}\Gamma_{2,2}\Gamma_{1,5}+21780\Gamma_{1,1}^{2}\Gamma_{2,2}\Gamma_{2,4}+18900\Gamma_{1,1}^{2}\Gamma_{2,2}\Gamma_{3,3}+23220\Gamma_{1,1}\Gamma_{1,3}^{3}+64440\Gamma_{1,1}\Gamma_{1,3}^{2}\Gamma_{2,2}\notag\\
&\qquad{}+61020\Gamma_{1,1}\Gamma_{1,3}\Gamma_{2,2}^{2}+18900\Gamma_{1,1}\Gamma_{2,2}^{3}+3\Gamma_{1,1}\Gamma_{4,6}+3\Gamma_{1,1}\Gamma_{5,5}+6\Gamma_{1,3}\Gamma_{2,6}+28\Gamma_{1,3}\Gamma_{3,5}+30\Gamma_{1,3}\Gamma_{4,4}\notag\\
&\qquad{}+30\Gamma_{2,2}\Gamma_{3,5}+30\Gamma_{2,2}\Gamma_{4,4}+3\Gamma_{1,5}^{2}+15\Gamma_{1,5}\Gamma_{2,4}+30\Gamma_{2,4}^{2}+60\Gamma_{2,4}\Gamma_{3,3}+30\Gamma_{3,3}^{2}\Bigr)\Biggr\},\notag\\[3pt]
V_{G^{11}}={}&-\Biggl\{\frac{m_1m_2^{11}+m_1^{11}m_2}{39916800}\Bigl(\Gamma_{1,1}^{11}+165\Gamma_{1,1}^{8}\Gamma_{1,3}+462\Gamma_{1,1}^{6}\Gamma_{1,5}+4620\Gamma_{1,1}^{5}\Gamma_{1,3}^{2}+330\Gamma_{1,1}^{4}\Gamma_{1,7}+9240\Gamma_{1,1}^{3}\Gamma_{1,3}\Gamma_{1,5}\label{meth:GammaG11}\\*
&\qquad{}+15400\Gamma_{1,1}^{2}\Gamma_{1,3}^{3}+55\Gamma_{1,1}^{2}\Gamma_{1,9}+1320\Gamma_{1,1}\Gamma_{1,3}\Gamma_{1,7}+1386\Gamma_{1,1}\Gamma_{1,5}^{2}+4620\Gamma_{1,3}^{2}\Gamma_{1,5}+\Gamma_{1,11}\Bigr)\notag\\
&\quad+\frac{m_1^{2}m_2^{10}+m_1^{10}m_2^{2}}{7257600}\Bigl(5120\Gamma_{1,1}^{11}+199680\Gamma_{1,1}^{8}\Gamma_{1,3}+11520\Gamma_{1,1}^{8}\Gamma_{2,2}+120960\Gamma_{1,1}^{6}\Gamma_{1,5}+13440\Gamma_{1,1}^{6}\Gamma_{2,4}\notag\\
&\qquad{}+1317120\Gamma_{1,1}^{5}\Gamma_{1,3}^{2}+161280\Gamma_{1,1}^{5}\Gamma_{1,3}\Gamma_{2,2}+16320\Gamma_{1,1}^{4}\Gamma_{1,7}+3360\Gamma_{1,1}^{4}\Gamma_{2,6}+564480\Gamma_{1,1}^{3}\Gamma_{1,3}\Gamma_{1,5}\notag\\
&\qquad{}+67200\Gamma_{1,1}^{3}\Gamma_{1,3}\Gamma_{2,4}+40320\Gamma_{1,1}^{3}\Gamma_{2,2}\Gamma_{1,5}+1030400\Gamma_{1,1}^{2}\Gamma_{1,3}^{3}+201600\Gamma_{1,1}^{2}\Gamma_{1,3}^{2}\Gamma_{2,2}+380\Gamma_{1,1}^{2}\Gamma_{1,9}\notag\\
&\qquad{}+180\Gamma_{1,1}^{2}\Gamma_{2,8}+14880\Gamma_{1,1}\Gamma_{1,3}\Gamma_{1,7}+3360\Gamma_{1,1}\Gamma_{1,3}\Gamma_{2,6}+1440\Gamma_{1,1}\Gamma_{2,2}\Gamma_{1,7}+17640\Gamma_{1,1}\Gamma_{1,5}^{2}\notag\\
&\qquad{}+5040\Gamma_{1,1}\Gamma_{1,5}\Gamma_{2,4}+65520\Gamma_{1,3}^{2}\Gamma_{1,5}+8400\Gamma_{1,3}^{2}\Gamma_{2,4}+10080\Gamma_{1,3}\Gamma_{2,2}\Gamma_{1,5}+\Gamma_{2,10}\Bigr)\notag\\
&\quad+\frac{m_1^{3}m_2^{9}+m_1^{9}m_2^{3}}{2177280}\Bigl(531441\Gamma_{1,1}^{11}+8325909\Gamma_{1,1}^{8}\Gamma_{1,3}+1574640\Gamma_{1,1}^{8}\Gamma_{2,2}+1745226\Gamma_{1,1}^{6}\Gamma_{1,5}+673596\Gamma_{1,1}^{6}\Gamma_{2,4}\notag\\
&\qquad{}+61236\Gamma_{1,1}^{6}\Gamma_{3,3}+21493836\Gamma_{1,1}^{5}\Gamma_{1,3}^{2}+9062928\Gamma_{1,1}^{5}\Gamma_{1,3}\Gamma_{2,2}+551124\Gamma_{1,1}^{5}\Gamma_{2,2}^{2}+67068\Gamma_{1,1}^{4}\Gamma_{1,7}+54432\Gamma_{1,1}^{4}\Gamma_{2,6}\notag\\
&\qquad{}+10206\Gamma_{1,1}^{4}\Gamma_{3,5}+2898504\Gamma_{1,1}^{3}\Gamma_{1,3}\Gamma_{1,5}+1360800\Gamma_{1,1}^{3}\Gamma_{1,3}\Gamma_{2,4}+136080\Gamma_{1,1}^{3}\Gamma_{1,3}\Gamma_{3,3}+816480\Gamma_{1,1}^{3}\Gamma_{2,2}\Gamma_{1,5}\notag\\
&\qquad{}+204120\Gamma_{1,1}^{3}\Gamma_{2,2}\Gamma_{2,4}+6327720\Gamma_{1,1}^{2}\Gamma_{1,3}^{3}+4626720\Gamma_{1,1}^{2}\Gamma_{1,3}^{2}\Gamma_{2,2}+612360\Gamma_{1,1}^{2}\Gamma_{1,3}\Gamma_{2,2}^{2}+243\Gamma_{1,1}^{2}\Gamma_{1,9}\notag\\
&\qquad{}+702\Gamma_{1,1}^{2}\Gamma_{2,8}+324\Gamma_{1,1}^{2}\Gamma_{3,7}+14256\Gamma_{1,1}\Gamma_{1,3}\Gamma_{1,7}+21168\Gamma_{1,1}\Gamma_{1,3}\Gamma_{2,6}+4536\Gamma_{1,1}\Gamma_{1,3}\Gamma_{3,5}+8208\Gamma_{1,1}\Gamma_{2,2}\Gamma_{1,7}\notag\\
&\qquad{}+4536\Gamma_{1,1}\Gamma_{2,2}\Gamma_{2,6}+17010\Gamma_{1,1}\Gamma_{1,5}^{2}+34776\Gamma_{1,1}\Gamma_{1,5}\Gamma_{2,4}+4536\Gamma_{1,1}\Gamma_{1,5}\Gamma_{3,3}+5670\Gamma_{1,1}\Gamma_{2,4}^{2}+106596\Gamma_{1,3}^{2}\Gamma_{1,5}\notag\\
&\qquad{}+68040\Gamma_{1,3}^{2}\Gamma_{2,4}+7560\Gamma_{1,3}^{2}\Gamma_{3,3}+81648\Gamma_{1,3}\Gamma_{2,2}\Gamma_{1,5}+22680\Gamma_{1,3}\Gamma_{2,2}\Gamma_{2,4}+6804\Gamma_{2,2}^{2}\Gamma_{1,5}+\Gamma_{3,9}\Bigr)\notag\\
&\quad+\frac{m_1^{4}m_2^{8}+m_1^{8}m_2^{4}}{967680}\Bigl(8388608\Gamma_{1,1}^{11}+65667072\Gamma_{1,1}^{8}\Gamma_{1,3}+27525120\Gamma_{1,1}^{8}\Gamma_{2,2}+5275648\Gamma_{1,1}^{6}\Gamma_{1,5}+5275648\Gamma_{1,1}^{6}\Gamma_{2,4}\notag\\
&\qquad{}+1548288\Gamma_{1,1}^{6}\Gamma_{3,3}+81543168\Gamma_{1,1}^{5}\Gamma_{1,3}^{2}+77758464\Gamma_{1,1}^{5}\Gamma_{1,3}\Gamma_{2,2}+13934592\Gamma_{1,1}^{5}\Gamma_{2,2}^{2}+59392\Gamma_{1,1}^{4}\Gamma_{1,7}\notag\\
&\qquad{}+150528\Gamma_{1,1}^{4}\Gamma_{2,6}+103936\Gamma_{1,1}^{4}\Gamma_{3,5}+17920\Gamma_{1,1}^{4}\Gamma_{4,4}+3540992\Gamma_{1,1}^{3}\Gamma_{1,3}\Gamma_{1,5}+4874240\Gamma_{1,1}^{3}\Gamma_{1,3}\Gamma_{2,4}\notag\\
&\qquad{}+1720320\Gamma_{1,1}^{3}\Gamma_{1,3}\Gamma_{3,3}+2580480\Gamma_{1,1}^{3}\Gamma_{2,2}\Gamma_{1,5}+2204160\Gamma_{1,1}^{3}\Gamma_{2,2}\Gamma_{2,4}+430080\Gamma_{1,1}^{3}\Gamma_{2,2}\Gamma_{3,3}\notag\\
&\qquad{}+10680320\Gamma_{1,1}^{2}\Gamma_{1,3}^{3}+18493440\Gamma_{1,1}^{2}\Gamma_{1,3}^{2}\Gamma_{2,2}+7741440\Gamma_{1,1}^{2}\Gamma_{1,3}\Gamma_{2,2}^{2}+645120\Gamma_{1,1}^{2}\Gamma_{2,2}^{3}+384\Gamma_{1,1}^{2}\Gamma_{2,8}\notag\\
&\qquad{}+992\Gamma_{1,1}^{2}\Gamma_{3,7}+448\Gamma_{1,1}^{2}\Gamma_{4,6}+2368\Gamma_{1,1}\Gamma_{1,3}\Gamma_{1,7}+18816\Gamma_{1,1}\Gamma_{1,3}\Gamma_{2,6}+21952\Gamma_{1,1}\Gamma_{1,3}\Gamma_{3,5}+4480\Gamma_{1,1}\Gamma_{1,3}\Gamma_{4,4}\notag\\
&\qquad{}+5376\Gamma_{1,1}\Gamma_{2,2}\Gamma_{1,7}+16128\Gamma_{1,1}\Gamma_{2,2}\Gamma_{2,6}+8064\Gamma_{1,1}\Gamma_{2,2}\Gamma_{3,5}+29568\Gamma_{1,1}\Gamma_{1,5}\Gamma_{2,4}+22848\Gamma_{1,1}\Gamma_{1,5}\Gamma_{3,3}\notag\\
&\qquad{}+23520\Gamma_{1,1}\Gamma_{2,4}^{2}+13440\Gamma_{1,1}\Gamma_{2,4}\Gamma_{3,3}+27328\Gamma_{1,3}^{2}\Gamma_{1,5}+98560\Gamma_{1,3}^{2}\Gamma_{2,4}+47040\Gamma_{1,3}^{2}\Gamma_{3,3}+88704\Gamma_{1,3}\Gamma_{2,2}\Gamma_{1,5}\notag\\
&\qquad{}+117600\Gamma_{1,3}\Gamma_{2,2}\Gamma_{2,4}+26880\Gamma_{1,3}\Gamma_{2,2}\Gamma_{3,3}+34272\Gamma_{2,2}^{2}\Gamma_{1,5}+20160\Gamma_{2,2}^{2}\Gamma_{2,4}+\Gamma_{4,8}\Bigr)\notag\\
&\quad+\frac{m_1^{5}m_2^{7}+m_1^{7}m_2^{5}}{604800}\Bigl(37515625\Gamma_{1,1}^{11}+177778125\Gamma_{1,1}^{8}\Gamma_{1,3}+128625000\Gamma_{1,1}^{8}\Gamma_{2,2}+4971400\Gamma_{1,1}^{6}\Gamma_{1,5}\notag\\
&\qquad{}+13116250\Gamma_{1,1}^{6}\Gamma_{2,4}+8268750\Gamma_{1,1}^{6}\Gamma_{3,3}+134011500\Gamma_{1,1}^{5}\Gamma_{1,3}^{2}+209055000\Gamma_{1,1}^{5}\Gamma_{1,3}\Gamma_{2,2}+74418750\Gamma_{1,1}^{5}\Gamma_{2,2}^{2}\notag\\
&\qquad{}+12005\Gamma_{1,1}^{4}\Gamma_{1,7}+109760\Gamma_{1,1}^{4}\Gamma_{2,6}+249725\Gamma_{1,1}^{4}\Gamma_{3,5}+140000\Gamma_{1,1}^{4}\Gamma_{4,4}+1954400\Gamma_{1,1}^{3}\Gamma_{1,3}\Gamma_{1,5}\notag\\
&\qquad{}+6851600\Gamma_{1,1}^{3}\Gamma_{1,3}\Gamma_{2,4}+4935000\Gamma_{1,1}^{3}\Gamma_{1,3}\Gamma_{3,3}+2058000\Gamma_{1,1}^{3}\Gamma_{2,2}\Gamma_{1,5}+5596500\Gamma_{1,1}^{3}\Gamma_{2,2}\Gamma_{2,4}\notag\\
&\qquad{}+3360000\Gamma_{1,1}^{3}\Gamma_{2,2}\Gamma_{3,3}+10935400\Gamma_{1,1}^{2}\Gamma_{1,3}^{3}+27812400\Gamma_{1,1}^{2}\Gamma_{1,3}^{2}\Gamma_{2,2}+22207500\Gamma_{1,1}^{2}\Gamma_{1,3}\Gamma_{2,2}^{2}\notag\\
&\qquad{}+5040000\Gamma_{1,1}^{2}\Gamma_{2,2}^{3}+490\Gamma_{1,1}^{2}\Gamma_{3,7}+1190\Gamma_{1,1}^{2}\Gamma_{4,6}+525\Gamma_{1,1}^{2}\Gamma_{5,5}+980\Gamma_{1,1}\Gamma_{1,3}\Gamma_{1,7}+6440\Gamma_{1,1}\Gamma_{1,3}\Gamma_{2,6}\notag\\
&\qquad{}+24500\Gamma_{1,1}\Gamma_{1,3}\Gamma_{3,5}+18200\Gamma_{1,1}\Gamma_{1,3}\Gamma_{4,4}+8820\Gamma_{1,1}\Gamma_{2,2}\Gamma_{2,6}+22680\Gamma_{1,1}\Gamma_{2,2}\Gamma_{3,5}+10500\Gamma_{1,1}\Gamma_{2,2}\Gamma_{4,4}\notag\\
&\qquad{}+1050\Gamma_{1,1}\Gamma_{1,5}^{2}+10500\Gamma_{1,1}\Gamma_{1,5}\Gamma_{2,4}+14700\Gamma_{1,1}\Gamma_{1,5}\Gamma_{3,3}+25200\Gamma_{1,1}\Gamma_{2,4}^{2}+42000\Gamma_{1,1}\Gamma_{2,4}\Gamma_{3,3}\notag\\
&\qquad{}+10500\Gamma_{1,1}\Gamma_{3,3}^{2}+19600\Gamma_{1,3}^{2}\Gamma_{1,5}+76300\Gamma_{1,3}^{2}\Gamma_{2,4}+65100\Gamma_{1,3}^{2}\Gamma_{3,3}+29400\Gamma_{1,3}\Gamma_{2,2}\Gamma_{1,5}+140700\Gamma_{1,3}\Gamma_{2,2}\Gamma_{2,4}\notag\\
&\qquad{}+109200\Gamma_{1,3}\Gamma_{2,2}\Gamma_{3,3}+22050\Gamma_{2,2}^{2}\Gamma_{1,5}+63000\Gamma_{2,2}^{2}\Gamma_{2,4}+31500\Gamma_{2,2}^{2}\Gamma_{3,3}+\Gamma_{5,7}\Bigr)\notag\\
&\quad+\frac{m_1^{6}m_2^{6}}{518400}\Bigl(60466176\Gamma_{1,1}^{11}+235146240\Gamma_{1,1}^{8}\Gamma_{1,3}+209952000\Gamma_{1,1}^{8}\Gamma_{2,2}+2892672\Gamma_{1,1}^{6}\Gamma_{1,5}+16329600\Gamma_{1,1}^{6}\Gamma_{2,4}\notag\\
&\qquad{}+13996800\Gamma_{1,1}^{6}\Gamma_{3,3}+150232320\Gamma_{1,1}^{5}\Gamma_{1,3}^{2}+270604800\Gamma_{1,1}^{5}\Gamma_{1,3}\Gamma_{2,2}+125971200\Gamma_{1,1}^{5}\Gamma_{2,2}^{2}\notag\\
&\qquad{}+38880\Gamma_{1,1}^{4}\Gamma_{2,6}+285120\Gamma_{1,1}^{4}\Gamma_{3,5}+259200\Gamma_{1,1}^{4}\Gamma_{4,4}+1347840\Gamma_{1,1}^{3}\Gamma_{1,3}\Gamma_{1,5}+7257600\Gamma_{1,1}^{3}\Gamma_{1,3}\Gamma_{2,4}\notag\\
&\qquad{}+6220800\Gamma_{1,1}^{3}\Gamma_{1,3}\Gamma_{3,3}+777600\Gamma_{1,1}^{3}\Gamma_{2,2}\Gamma_{1,5}+6609600\Gamma_{1,1}^{3}\Gamma_{2,2}\Gamma_{2,4}+6220800\Gamma_{1,1}^{3}\Gamma_{2,2}\Gamma_{3,3}\notag\\
&\qquad{}+10886400\Gamma_{1,1}^{2}\Gamma_{1,3}^{3}+29808000\Gamma_{1,1}^{2}\Gamma_{1,3}^{2}\Gamma_{2,2}+27993600\Gamma_{1,1}^{2}\Gamma_{1,3}\Gamma_{2,2}^{2}+9331200\Gamma_{1,1}^{2}\Gamma_{2,2}^{3}+1080\Gamma_{1,1}^{2}\Gamma_{4,6}\notag\\
&\qquad{}+1260\Gamma_{1,1}^{2}\Gamma_{5,5}+4320\Gamma_{1,1}\Gamma_{1,3}\Gamma_{2,6}+24480\Gamma_{1,1}\Gamma_{1,3}\Gamma_{3,5}+21600\Gamma_{1,1}\Gamma_{1,3}\Gamma_{4,4}+21600\Gamma_{1,1}\Gamma_{2,2}\Gamma_{3,5}\notag\\
&\qquad{}+25200\Gamma_{1,1}\Gamma_{2,2}\Gamma_{4,4}+3096\Gamma_{1,1}\Gamma_{1,5}^{2}+10800\Gamma_{1,1}\Gamma_{1,5}\Gamma_{2,4}+26100\Gamma_{1,1}\Gamma_{2,4}^{2}+43200\Gamma_{1,1}\Gamma_{2,4}\Gamma_{3,3}\notag\\
&\qquad{}+25200\Gamma_{1,1}\Gamma_{3,3}^{2}+15120\Gamma_{1,3}^{2}\Gamma_{1,5}+75600\Gamma_{1,3}^{2}\Gamma_{2,4}+64800\Gamma_{1,3}^{2}\Gamma_{3,3}+21600\Gamma_{1,3}\Gamma_{2,2}\Gamma_{1,5}\notag\\
&\qquad{}+140400\Gamma_{1,3}\Gamma_{2,2}\Gamma_{2,4}+129600\Gamma_{1,3}\Gamma_{2,2}\Gamma_{3,3}+64800\Gamma_{2,2}^{2}\Gamma_{2,4}+75600\Gamma_{2,2}^{2}\Gamma_{3,3}+\Gamma_{6,6}\Bigr)\Biggr\},\notag\\[3pt]
V_{G^{12}}={}&\Biggl\{\frac{m_1m_2^{12}+m_1^{12}m_2}{479001600}\Bigl(\Gamma_{1,1}^{12}+220\Gamma_{1,1}^{9}\Gamma_{1,3}+792\Gamma_{1,1}^{7}\Gamma_{1,5}+9240\Gamma_{1,1}^{6}\Gamma_{1,3}^{2}+792\Gamma_{1,1}^{5}\Gamma_{1,7}\label{meth:GammaG12}\\*
&\qquad{}+27720\Gamma_{1,1}^{4}\Gamma_{1,3}\Gamma_{1,5}+61600\Gamma_{1,1}^{3}\Gamma_{1,3}^{3}+220\Gamma_{1,1}^{3}\Gamma_{1,9}+7920\Gamma_{1,1}^{2}\Gamma_{1,3}\Gamma_{1,7}+8316\Gamma_{1,1}^{2}\Gamma_{1,5}^{2}\notag\\
&\qquad{}+55440\Gamma_{1,1}\Gamma_{1,3}^{2}\Gamma_{1,5}+15400\Gamma_{1,3}^{4}+12\Gamma_{1,1}\Gamma_{1,11}+220\Gamma_{1,3}\Gamma_{1,9}+792\Gamma_{1,5}\Gamma_{1,7}\Bigr)\notag\\
&\quad+\frac{m_1^{2}m_2^{11}+m_1^{11}m_2^{2}}{3628800}\Bigl(512\Gamma_{1,1}^{12}+26880\Gamma_{1,1}^{9}\Gamma_{1,3}+1280\Gamma_{1,1}^{9}\Gamma_{2,2}+21504\Gamma_{1,1}^{7}\Gamma_{1,5}+1920\Gamma_{1,1}^{7}\Gamma_{2,4}\notag\\
&\qquad{}+268800\Gamma_{1,1}^{6}\Gamma_{1,3}^{2}+26880\Gamma_{1,1}^{6}\Gamma_{1,3}\Gamma_{2,2}+4320\Gamma_{1,1}^{5}\Gamma_{1,7}+672\Gamma_{1,1}^{5}\Gamma_{2,6}+178080\Gamma_{1,1}^{4}\Gamma_{1,3}\Gamma_{1,5}\notag\\
&\qquad{}+16800\Gamma_{1,1}^{4}\Gamma_{1,3}\Gamma_{2,4}+10080\Gamma_{1,1}^{4}\Gamma_{2,2}\Gamma_{1,5}+425600\Gamma_{1,1}^{3}\Gamma_{1,3}^{3}+67200\Gamma_{1,1}^{3}\Gamma_{1,3}^{2}\Gamma_{2,2}+200\Gamma_{1,1}^{3}\Gamma_{1,9}+60\Gamma_{1,1}^{3}\Gamma_{2,8}\notag\\
&\qquad{}+10080\Gamma_{1,1}^{2}\Gamma_{1,3}\Gamma_{1,7}+1680\Gamma_{1,1}^{2}\Gamma_{1,3}\Gamma_{2,6}+720\Gamma_{1,1}^{2}\Gamma_{2,2}\Gamma_{1,7}+11592\Gamma_{1,1}^{2}\Gamma_{1,5}^{2}+2520\Gamma_{1,1}^{2}\Gamma_{1,5}\Gamma_{2,4}\notag\\
&\qquad{}+84000\Gamma_{1,1}\Gamma_{1,3}^{2}\Gamma_{1,5}+8400\Gamma_{1,1}\Gamma_{1,3}^{2}\Gamma_{2,4}+10080\Gamma_{1,1}\Gamma_{1,3}\Gamma_{2,2}\Gamma_{1,5}+25200\Gamma_{1,3}^{4}+5600\Gamma_{1,3}^{3}\Gamma_{2,2}+\Gamma_{1,1}\Gamma_{1,11}\notag\\
&\qquad{}+\Gamma_{1,1}\Gamma_{2,10}+45\Gamma_{1,3}\Gamma_{1,9}+15\Gamma_{1,3}\Gamma_{2,8}+5\Gamma_{2,2}\Gamma_{1,9}+210\Gamma_{1,5}\Gamma_{1,7}+42\Gamma_{1,5}\Gamma_{2,6}+30\Gamma_{2,4}\Gamma_{1,7}\Bigr)\notag\\
&\quad+\frac{m_1^{3}m_2^{10}+m_1^{10}m_2^{3}}{3628800}\Bigl(328050\Gamma_{1,1}^{12}+7031205\Gamma_{1,1}^{9}\Gamma_{1,3}+1082565\Gamma_{1,1}^{9}\Gamma_{2,2}+2041200\Gamma_{1,1}^{7}\Gamma_{1,5}+612360\Gamma_{1,1}^{7}\Gamma_{2,4}\notag\\
&\qquad{}+43740\Gamma_{1,1}^{7}\Gamma_{3,3}+28202580\Gamma_{1,1}^{6}\Gamma_{1,3}^{2}+9491580\Gamma_{1,1}^{6}\Gamma_{1,3}\Gamma_{2,2}+459270\Gamma_{1,1}^{6}\Gamma_{2,2}^{2}+129600\Gamma_{1,1}^{5}\Gamma_{1,7}\notag\\
&\qquad{}+73710\Gamma_{1,1}^{5}\Gamma_{2,6}+10206\Gamma_{1,1}^{5}\Gamma_{3,5}+6401430\Gamma_{1,1}^{4}\Gamma_{1,3}\Gamma_{1,5}+2211300\Gamma_{1,1}^{4}\Gamma_{1,3}\Gamma_{2,4}+170100\Gamma_{1,1}^{4}\Gamma_{1,3}\Gamma_{3,3}\notag\\
&\qquad{}+1343790\Gamma_{1,1}^{4}\Gamma_{2,2}\Gamma_{1,5}+255150\Gamma_{1,1}^{4}\Gamma_{2,2}\Gamma_{2,4}+17476200\Gamma_{1,1}^{3}\Gamma_{1,3}^{3}+9865800\Gamma_{1,1}^{3}\Gamma_{1,3}^{2}\Gamma_{2,2}\notag\\
&\qquad{}+1020600\Gamma_{1,1}^{3}\Gamma_{1,3}\Gamma_{2,2}^{2}+1400\Gamma_{1,1}^{3}\Gamma_{1,9}+1890\Gamma_{1,1}^{3}\Gamma_{2,8}+540\Gamma_{1,1}^{3}\Gamma_{3,7}+95220\Gamma_{1,1}^{2}\Gamma_{1,3}\Gamma_{1,7}+74340\Gamma_{1,1}^{2}\Gamma_{1,3}\Gamma_{2,6}\notag\\
&\qquad{}+11340\Gamma_{1,1}^{2}\Gamma_{1,3}\Gamma_{3,5}+30780\Gamma_{1,1}^{2}\Gamma_{2,2}\Gamma_{1,7}+11340\Gamma_{1,1}^{2}\Gamma_{2,2}\Gamma_{2,6}+113400\Gamma_{1,1}^{2}\Gamma_{1,5}^{2}+120960\Gamma_{1,1}^{2}\Gamma_{1,5}\Gamma_{2,4}\notag\\
&\qquad{}+11340\Gamma_{1,1}^{2}\Gamma_{1,5}\Gamma_{3,3}+14175\Gamma_{1,1}^{2}\Gamma_{2,4}^{2}+1081080\Gamma_{1,1}\Gamma_{1,3}^{2}\Gamma_{1,5}+453600\Gamma_{1,1}\Gamma_{1,3}^{2}\Gamma_{2,4}+37800\Gamma_{1,1}\Gamma_{1,3}^{2}\Gamma_{3,3}\notag\\
&\qquad{}+551880\Gamma_{1,1}\Gamma_{1,3}\Gamma_{2,2}\Gamma_{1,5}+113400\Gamma_{1,1}\Gamma_{1,3}\Gamma_{2,2}\Gamma_{2,4}+34020\Gamma_{1,1}\Gamma_{2,2}^{2}\Gamma_{1,5}+390600\Gamma_{1,3}^{4}+340200\Gamma_{1,3}^{3}\Gamma_{2,2}\notag\\
&\qquad{}+56700\Gamma_{1,3}^{2}\Gamma_{2,2}^{2}+5\Gamma_{1,1}\Gamma_{2,10}+5\Gamma_{1,1}\Gamma_{3,9}+5\Gamma_{1,3}\Gamma_{1,9}+180\Gamma_{1,3}\Gamma_{2,8}+60\Gamma_{1,3}\Gamma_{3,7}+45\Gamma_{2,2}\Gamma_{1,9}\notag\\
&\qquad{}+45\Gamma_{2,2}\Gamma_{2,8}+630\Gamma_{1,5}\Gamma_{2,6}+126\Gamma_{1,5}\Gamma_{3,5}+420\Gamma_{2,4}\Gamma_{1,7}+210\Gamma_{2,4}\Gamma_{2,6}+60\Gamma_{3,3}\Gamma_{1,7}\Bigr)\notag\\
&\quad+\frac{m_1^{4}m_2^{9}+m_1^{9}m_2^{4}}{725760}\Bigl(3981312\Gamma_{1,1}^{12}+43180032\Gamma_{1,1}^{9}\Gamma_{1,3}+14598144\Gamma_{1,1}^{9}\Gamma_{2,2}+5225472\Gamma_{1,1}^{7}\Gamma_{1,5}+3823104\Gamma_{1,1}^{7}\Gamma_{2,4}\notag\\
&\qquad{}+860160\Gamma_{1,1}^{7}\Gamma_{3,3}+85091328\Gamma_{1,1}^{6}\Gamma_{1,3}^{2}+64318464\Gamma_{1,1}^{6}\Gamma_{1,3}\Gamma_{2,2}+9031680\Gamma_{1,1}^{6}\Gamma_{2,2}^{2}+116640\Gamma_{1,1}^{5}\Gamma_{1,7}\notag\\
&\qquad{}+181440\Gamma_{1,1}^{5}\Gamma_{2,6}+86016\Gamma_{1,1}^{5}\Gamma_{3,5}+10752\Gamma_{1,1}^{5}\Gamma_{4,4}+7211232\Gamma_{1,1}^{4}\Gamma_{1,3}\Gamma_{1,5}+6619200\Gamma_{1,1}^{4}\Gamma_{1,3}\Gamma_{2,4}\notag\\
&\qquad{}+1720320\Gamma_{1,1}^{4}\Gamma_{1,3}\Gamma_{3,3}+3737664\Gamma_{1,1}^{4}\Gamma_{2,2}\Gamma_{1,5}+2257920\Gamma_{1,1}^{4}\Gamma_{2,2}\Gamma_{2,4}+322560\Gamma_{1,1}^{4}\Gamma_{2,2}\Gamma_{3,3}+24554880\Gamma_{1,1}^{3}\Gamma_{1,3}^{3}\notag\\
&\qquad{}+32497920\Gamma_{1,1}^{3}\Gamma_{1,3}^{2}\Gamma_{2,2}+10321920\Gamma_{1,1}^{3}\Gamma_{1,3}\Gamma_{2,2}^{2}+645120\Gamma_{1,1}^{3}\Gamma_{2,2}^{3}+243\Gamma_{1,1}^{3}\Gamma_{1,9}+1377\Gamma_{1,1}^{3}\Gamma_{2,8}\notag\\
&\qquad{}+1632\Gamma_{1,1}^{3}\Gamma_{3,7}+448\Gamma_{1,1}^{3}\Gamma_{4,6}+26460\Gamma_{1,1}^{2}\Gamma_{1,3}\Gamma_{1,7}+74340\Gamma_{1,1}^{2}\Gamma_{1,3}\Gamma_{2,6}+47712\Gamma_{1,1}^{2}\Gamma_{1,3}\Gamma_{3,5}+6720\Gamma_{1,1}^{2}\Gamma_{1,3}\Gamma_{4,4}\notag\\
&\qquad{}+26244\Gamma_{1,1}^{2}\Gamma_{2,2}\Gamma_{1,7}+38304\Gamma_{1,1}^{2}\Gamma_{2,2}\Gamma_{2,6}+12096\Gamma_{1,1}^{2}\Gamma_{2,2}\Gamma_{3,5}+25515\Gamma_{1,1}^{2}\Gamma_{1,5}^{2}+120834\Gamma_{1,1}^{2}\Gamma_{1,5}\Gamma_{2,4}\notag\\
&\qquad{}+50400\Gamma_{1,1}^{2}\Gamma_{1,5}\Gamma_{3,3}+52920\Gamma_{1,1}^{2}\Gamma_{2,4}^{2}+20160\Gamma_{1,1}^{2}\Gamma_{2,4}\Gamma_{3,3}+423108\Gamma_{1,1}\Gamma_{1,3}^{2}\Gamma_{1,5}+609420\Gamma_{1,1}\Gamma_{1,3}^{2}\Gamma_{2,4}\notag\\
&\qquad{}+194880\Gamma_{1,1}\Gamma_{1,3}^{2}\Gamma_{3,3}+645624\Gamma_{1,1}\Gamma_{1,3}\Gamma_{2,2}\Gamma_{1,5}+504000\Gamma_{1,1}\Gamma_{1,3}\Gamma_{2,2}\Gamma_{2,4}+80640\Gamma_{1,1}\Gamma_{1,3}\Gamma_{2,2}\Gamma_{3,3}\notag\\
&\qquad{}+151200\Gamma_{1,1}\Gamma_{2,2}^{2}\Gamma_{1,5}+60480\Gamma_{1,1}\Gamma_{2,2}^{2}\Gamma_{2,4}+227010\Gamma_{1,3}^{4}+516600\Gamma_{1,3}^{3}\Gamma_{2,2}+292320\Gamma_{1,3}^{2}\Gamma_{2,2}^{2}+40320\Gamma_{1,3}\Gamma_{2,2}^{3}\notag\\
&\qquad{}+3\Gamma_{1,1}\Gamma_{3,9}+3\Gamma_{1,1}\Gamma_{4,8}+3\Gamma_{1,3}\Gamma_{1,9}+9\Gamma_{1,3}\Gamma_{2,8}+84\Gamma_{1,3}\Gamma_{3,7}+28\Gamma_{1,3}\Gamma_{4,6}+36\Gamma_{2,2}\Gamma_{2,8}+36\Gamma_{2,2}\Gamma_{3,7}\notag\\
&\qquad{}+210\Gamma_{1,5}\Gamma_{3,5}+42\Gamma_{1,5}\Gamma_{4,4}+12\Gamma_{2,4}\Gamma_{1,7}+252\Gamma_{2,4}\Gamma_{2,6}+126\Gamma_{2,4}\Gamma_{3,5}+84\Gamma_{3,3}\Gamma_{1,7}+84\Gamma_{3,3}\Gamma_{2,6}\Bigr)\notag\\
&\quad+\frac{m_1^{5}m_2^{8}+m_1^{8}m_2^{5}}{120960}\Bigl(8000000\Gamma_{1,1}^{12}+51300000\Gamma_{1,1}^{9}\Gamma_{1,3}+30800000\Gamma_{1,1}^{9}\Gamma_{2,2}+2524425\Gamma_{1,1}^{7}\Gamma_{1,5}+4291875\Gamma_{1,1}^{7}\Gamma_{2,4}\notag\\
&\qquad{}+2100000\Gamma_{1,1}^{7}\Gamma_{3,3}+59044125\Gamma_{1,1}^{6}\Gamma_{1,3}^{2}+77568750\Gamma_{1,1}^{6}\Gamma_{1,3}\Gamma_{2,2}+22050000\Gamma_{1,1}^{6}\Gamma_{2,2}^{2}+18432\Gamma_{1,1}^{5}\Gamma_{1,7}\notag\\
&\qquad{}+76160\Gamma_{1,1}^{5}\Gamma_{2,6}+99575\Gamma_{1,1}^{5}\Gamma_{3,5}+39375\Gamma_{1,1}^{5}\Gamma_{4,4}+1850835\Gamma_{1,1}^{4}\Gamma_{1,3}\Gamma_{1,5}+4100775\Gamma_{1,1}^{4}\Gamma_{1,3}\Gamma_{2,4}\notag\\
&\qquad{}+2315250\Gamma_{1,1}^{4}\Gamma_{1,3}\Gamma_{3,3}+1706880\Gamma_{1,1}^{4}\Gamma_{2,2}\Gamma_{1,5}+2735250\Gamma_{1,1}^{4}\Gamma_{2,2}\Gamma_{2,4}+1181250\Gamma_{1,1}^{4}\Gamma_{2,2}\Gamma_{3,3}\notag\\
&\qquad{}+9827300\Gamma_{1,1}^{3}\Gamma_{1,3}^{3}+21642600\Gamma_{1,1}^{3}\Gamma_{1,3}^{2}\Gamma_{2,2}+13891500\Gamma_{1,1}^{3}\Gamma_{1,3}\Gamma_{2,2}^{2}+2362500\Gamma_{1,1}^{3}\Gamma_{2,2}^{3}+128\Gamma_{1,1}^{3}\Gamma_{2,8}\notag\\
&\qquad{}+608\Gamma_{1,1}^{3}\Gamma_{3,7}+665\Gamma_{1,1}^{3}\Gamma_{4,6}+175\Gamma_{1,1}^{3}\Gamma_{5,5}+2208\Gamma_{1,1}^{2}\Gamma_{1,3}\Gamma_{1,7}+13874\Gamma_{1,1}^{2}\Gamma_{1,3}\Gamma_{2,6}+27524\Gamma_{1,1}^{2}\Gamma_{1,3}\Gamma_{3,5}\notag\\
&\qquad{}+13650\Gamma_{1,1}^{2}\Gamma_{1,3}\Gamma_{4,4}+2688\Gamma_{1,1}^{2}\Gamma_{2,2}\Gamma_{1,7}+15456\Gamma_{1,1}^{2}\Gamma_{2,2}\Gamma_{2,6}+18690\Gamma_{1,1}^{2}\Gamma_{2,2}\Gamma_{3,5}+5250\Gamma_{1,1}^{2}\Gamma_{2,2}\Gamma_{4,4}\notag\\
&\qquad{}+525\Gamma_{1,1}^{2}\Gamma_{1,5}^{2}+21735\Gamma_{1,1}^{2}\Gamma_{1,5}\Gamma_{2,4}+23520\Gamma_{1,1}^{2}\Gamma_{1,5}\Gamma_{3,3}+28770\Gamma_{1,1}^{2}\Gamma_{2,4}^{2}+33600\Gamma_{1,1}^{2}\Gamma_{2,4}\Gamma_{3,3}\notag\\
&\qquad{}+5250\Gamma_{1,1}^{2}\Gamma_{3,3}^{2}+60900\Gamma_{1,1}\Gamma_{1,3}^{2}\Gamma_{1,5}+195090\Gamma_{1,1}\Gamma_{1,3}^{2}\Gamma_{2,4}+135240\Gamma_{1,1}\Gamma_{1,3}^{2}\Gamma_{3,3}+128100\Gamma_{1,1}\Gamma_{1,3}\Gamma_{2,2}\Gamma_{1,5}\notag\\
&\qquad{}+307860\Gamma_{1,1}\Gamma_{1,3}\Gamma_{2,2}\Gamma_{2,4}+163800\Gamma_{1,1}\Gamma_{1,3}\Gamma_{2,2}\Gamma_{3,3}+70560\Gamma_{1,1}\Gamma_{2,2}^{2}\Gamma_{1,5}+100800\Gamma_{1,1}\Gamma_{2,2}^{2}\Gamma_{2,4}\notag\\
&\qquad{}+31500\Gamma_{1,1}\Gamma_{2,2}^{2}\Gamma_{3,3}+52290\Gamma_{1,3}^{4}+176540\Gamma_{1,3}^{3}\Gamma_{2,2}+202860\Gamma_{1,3}^{2}\Gamma_{2,2}^{2}+81900\Gamma_{1,3}\Gamma_{2,2}^{3}+7875\Gamma_{2,2}^{4}+\Gamma_{1,1}\Gamma_{4,8}\notag\\
&\qquad{}+\Gamma_{1,1}\Gamma_{5,7}+2\Gamma_{1,3}\Gamma_{2,8}+6\Gamma_{1,3}\Gamma_{3,7}+21\Gamma_{1,3}\Gamma_{4,6}+7\Gamma_{1,3}\Gamma_{5,5}+14\Gamma_{2,2}\Gamma_{3,7}+14\Gamma_{2,2}\Gamma_{4,6}+\Gamma_{1,5}\Gamma_{1,7}+7\Gamma_{1,5}\Gamma_{3,5}\notag\\
&\qquad{}+35\Gamma_{1,5}\Gamma_{4,4}+7\Gamma_{2,4}\Gamma_{1,7}+14\Gamma_{2,4}\Gamma_{2,6}+70\Gamma_{2,4}\Gamma_{3,5}+35\Gamma_{2,4}\Gamma_{4,4}+42\Gamma_{3,3}\Gamma_{2,6}+42\Gamma_{3,3}\Gamma_{3,5}\Bigr)\notag\\
&\quad+\frac{m_1^{6}m_2^{7}+m_1^{7}m_2^{6}}{259200}\Bigl(56010528\Gamma_{1,1}^{12}+262589040\Gamma_{1,1}^{9}\Gamma_{1,3}+220041360\Gamma_{1,1}^{9}\Gamma_{2,2}+4988736\Gamma_{1,1}^{7}\Gamma_{1,5}\notag\\
&\qquad{}+20423340\Gamma_{1,1}^{7}\Gamma_{2,4}+15876000\Gamma_{1,1}^{7}\Gamma_{3,3}+227198160\Gamma_{1,1}^{6}\Gamma_{1,3}^{2}+389347560\Gamma_{1,1}^{6}\Gamma_{1,3}\Gamma_{2,2}+166698000\Gamma_{1,1}^{6}\Gamma_{2,2}^{2}\notag\\
&\qquad{}+7203\Gamma_{1,1}^{5}\Gamma_{1,7}+109419\Gamma_{1,1}^{5}\Gamma_{2,6}+440640\Gamma_{1,1}^{5}\Gamma_{3,5}+340200\Gamma_{1,1}^{5}\Gamma_{4,4}+3270465\Gamma_{1,1}^{4}\Gamma_{1,3}\Gamma_{1,5}\notag\\
&\qquad{}+14394375\Gamma_{1,1}^{4}\Gamma_{1,3}\Gamma_{2,4}+11548800\Gamma_{1,1}^{4}\Gamma_{1,3}\Gamma_{3,3}+2622285\Gamma_{1,1}^{4}\Gamma_{2,2}\Gamma_{1,5}+12592800\Gamma_{1,1}^{4}\Gamma_{2,2}\Gamma_{2,4}\notag\\
&\qquad{}+10206000\Gamma_{1,1}^{4}\Gamma_{2,2}\Gamma_{3,3}+29823300\Gamma_{1,1}^{3}\Gamma_{1,3}^{3}+78399900\Gamma_{1,1}^{3}\Gamma_{1,3}^{2}\Gamma_{2,2}+69292800\Gamma_{1,1}^{3}\Gamma_{1,3}\Gamma_{2,2}^{2}\notag\\
&\qquad{}+20412000\Gamma_{1,1}^{3}\Gamma_{2,2}^{3}+490\Gamma_{1,1}^{3}\Gamma_{3,7}+2640\Gamma_{1,1}^{3}\Gamma_{4,6}+2160\Gamma_{1,1}^{3}\Gamma_{5,5}+1470\Gamma_{1,1}^{2}\Gamma_{1,3}\Gamma_{1,7}+19410\Gamma_{1,1}^{2}\Gamma_{1,3}\Gamma_{2,6}\notag\\
&\qquad{}+85410\Gamma_{1,1}^{2}\Gamma_{1,3}\Gamma_{3,5}+69750\Gamma_{1,1}^{2}\Gamma_{1,3}\Gamma_{4,4}+13230\Gamma_{1,1}^{2}\Gamma_{2,2}\Gamma_{2,6}+77310\Gamma_{1,1}^{2}\Gamma_{2,2}\Gamma_{3,5}+64800\Gamma_{1,1}^{2}\Gamma_{2,2}\Gamma_{4,4}\notag\\
&\qquad{}+8208\Gamma_{1,1}^{2}\Gamma_{1,5}^{2}+38340\Gamma_{1,1}^{2}\Gamma_{1,5}\Gamma_{2,4}+22050\Gamma_{1,1}^{2}\Gamma_{1,5}\Gamma_{3,3}+89550\Gamma_{1,1}^{2}\Gamma_{2,4}^{2}+148950\Gamma_{1,1}^{2}\Gamma_{2,4}\Gamma_{3,3}\notag\\
&\qquad{}+64800\Gamma_{1,1}^{2}\Gamma_{3,3}^{2}+121860\Gamma_{1,1}\Gamma_{1,3}^{2}\Gamma_{1,5}+548100\Gamma_{1,1}\Gamma_{1,3}^{2}\Gamma_{2,4}+449100\Gamma_{1,1}\Gamma_{1,3}^{2}\Gamma_{3,3}+184860\Gamma_{1,1}\Gamma_{1,3}\Gamma_{2,2}\Gamma_{1,5}\notag\\
&\qquad{}+977400\Gamma_{1,1}\Gamma_{1,3}\Gamma_{2,2}\Gamma_{2,4}+837000\Gamma_{1,1}\Gamma_{1,3}\Gamma_{2,2}\Gamma_{3,3}+66150\Gamma_{1,1}\Gamma_{2,2}^{2}\Gamma_{1,5}+446850\Gamma_{1,1}\Gamma_{2,2}^{2}\Gamma_{2,4}\notag\\
&\qquad{}+388800\Gamma_{1,1}\Gamma_{2,2}^{2}\Gamma_{3,3}+137700\Gamma_{1,3}^{4}+494100\Gamma_{1,3}^{3}\Gamma_{2,2}+673650\Gamma_{1,3}^{2}\Gamma_{2,2}^{2}+418500\Gamma_{1,3}\Gamma_{2,2}^{3}+97200\Gamma_{2,2}^{4}\notag\\
&\qquad{}+3\Gamma_{1,1}\Gamma_{5,7}+3\Gamma_{1,1}\Gamma_{6,6}+10\Gamma_{1,3}\Gamma_{3,7}+45\Gamma_{1,3}\Gamma_{4,6}+45\Gamma_{1,3}\Gamma_{5,5}+45\Gamma_{2,2}\Gamma_{4,6}+45\Gamma_{2,2}\Gamma_{5,5}+3\Gamma_{1,5}\Gamma_{1,7}\notag\\
&\qquad{}+24\Gamma_{1,5}\Gamma_{2,6}+45\Gamma_{1,5}\Gamma_{3,5}+45\Gamma_{2,4}\Gamma_{2,6}+165\Gamma_{2,4}\Gamma_{3,5}+150\Gamma_{2,4}\Gamma_{4,4}+150\Gamma_{3,3}\Gamma_{3,5}+150\Gamma_{3,3}\Gamma_{4,4}\Bigr)\Biggr\}.\notag
\end{align}
\endgroup

\newpage
\phantomsection\pdfbookmark[1]{References}{references}
\input{ref.bbl}
\end{document}

%% file: ref.bbl
\providecommand{\href}[2]{#2}\begingroup\raggedright\endgroup